\documentclass{aa}
\usepackage{multirow}
\usepackage{booktabs}
\usepackage[varg]{txfonts}
\usepackage{graphicx}	
\usepackage{amsmath}	
\usepackage{subfigure}
\usepackage{epstopdf}
\usepackage[colorlinks, linkcolor=blue, urlcolor=blue, citecolor=blue]{hyperref}
\usepackage{orcidlink}

\begin{document}
	
	\title{R-process nucleosynthesis from magnetar giant flares in neutron star--white dwarf mergers: A unified picture for peculiar long gamma-ray bursts}

	\author{Shu-Qing Zhong\orcidlink{0000-0002-1766-6947}\inst{\ref{inst1}}
	\and Long Li\orcidlink{0000-0002-8391-5980}\inst{\ref{inst2}}
	\and Le Zou\orcidlink{0000-0003-4639-5397}\inst{\ref{inst3},\ref{inst4}}
	\and Ji-Gui Cheng\orcidlink{0000-0002-2585-442X}\inst{\ref{inst5}}
	\and Yan-Zhi Meng\orcidlink{0000-0002-1122-1146}\inst{\ref{inst1}}
	\and Jia-Hong Gu\inst{\ref{inst1}} 
	}
	
	\institute{School of Science, Guangxi University of Science and Technology, Liuzhou 545006, PR China\label{inst1}; \email{\href{mailto:sq_zhong@qq.com}{sq\_zhong@qq.com}}
		\and Department of Physics, School of Physics and Materials Science, Nanchang University, Nanchang 330031, PR China\label{inst2}
		\and Department of Physics, Xiangtan University, Xiangtan 411105, PR China\label{inst3}
		\and Key Laboratory of Stars and Interstellar Medium, Xiangtan University, Xiangtan 411105, PR China\label{inst4}
		\and School of Physics and Electronics, Hunan University of Science and Technology, Xiangtan 411201, PR China\label{inst5}
		}

	\abstract{Peculiar long gamma-ray bursts (GRBs), exemplified by GRBs 211211A and 230307A, exhibit a long-duration multi-component prompt emission, an X-ray plateau in their afterglow, and a kilonova signature. Their origin remains highly debated. In this work, we present a unified picture for these events based on neutron star--white dwarf (NS--WD) mergers involving a pre-merger magnetar and a massive WD. In this picture, tidal disruption of the WD forms a constant-entropy accretion disk. Hyperaccretion from this disk onto the NS during the early accretion phase amplifies its toroidal magnetic field to strengths sufficient to trigger repeated magnetar giant flares (GFs). The main burst (MB) of the prompt emission consists of a ``forest'' of initial spikes from these GFs, while the subsequent magnetic propeller phase generates the extended emission (EE) and naturally explains the observed MB--EE trough. Crucially, the $e^{\pm}$-$\gamma$ fireball associated with each GF initial spike shocks the NS crust, leading to crustal ejection that synthesizes r-process heavy elements via the $\alpha$-rich freeze-out mechanism, thereby resolving the r-process deficit in conventional NS--WD hydrodynamic simulations. The ensemble of such fireballs over the MB duration collectively yields $M_{\rm ej}\gtrsim 10^{-5}-10^{-3}\,M_\odot$ of ejecta, sufficient to power the observed kilonova signature when further boosted by the spin-down of the post-merger magnetar. Meanwhile, the spin-down radiation also powers the X-ray plateau. This tidally disrupted NS--WD merger picture provides a self-consistent framework that unifies the prompt emission, afterglow, kilonova, and r-process nucleosynthesis observed in peculiar long GRBs.}
	
	\titlerunning{Unified picture for peculiar LGRBs}
	\authorrunning{Zhong, S.-Q., et al.}
	\maketitle

\section{Introduction}\label{sec:intro} 
Traditionally, gamma-ray bursts (GRBs) are classified into long GRBs (LGRBs) and short GRBs (SGRBs) based on the observed bimodal duration distribution \citep{kou93}. LGRBs are generally believed to originate from the core collapse of massive stars, as supported by associations with supernovae in several cases \citep[e.g.,][]{nara92,woo93,gala98,hior03,stan03}. 
In contrast, SGRBs are typically attributed to mergers of binary compact objects \citep{pacz86,pacz91,eich89}, a paradigm strongly confirmed by the detection of SGRB 170817A in coincidence with the gravitational-wave event GW170817 and the kilonova AT 2017gfo from a neutron star–neutron star (NS–NS) merger \citep[e.g.,][]{abbott17c,abbott17a,abbott17b,cou17,tro17}.

However, a number of peculiar LGRBs, with durations exceeding 2 s but exhibiting other properties characteristic of SGRBs, challenge this standard classification. Examples include GRB 060614 \citep{gal06,della06,fyn06,geh06}, GRB 211227A \citep{lv22}, GRB 211211A \citep{ras22,yang22,tro22,zhang22,mei22}, and GRB 230307A \citep{levan24,sun25,gill23,yang24}. These events are increasingly interpreted as arising from compact binary mergers, though the specific progenitor systems remain under debate.

Proposed models fall into two general classes. The first adopts traditional SGRB progenitors but introduces additional physical mechanisms to prolong the emission duration. For GRB 211211A, such models include a neutron star–black hole (NS–BH) merger with ejecta fallback accretion \citep{zhu22} or a magnetic barrier \citep{gao22} or a photosphere with a structured jet \citep{meng24}, or an NS–NS/BH merger leaving behind a BH engine surrounded by a massive accretion disk \citep{gott23}. The second class invokes non-traditional progenitors, such as the accretion-induced collapse (AIC) of a strongly magnetized and rapidly rotating white dwarf (WD) \citep{cheong25}, or an NS–WD merger involving a magnetar and a massive WD \citep{yang22,zhong23,wang24}. 
Some studies further suggest that GRBs 211211A and 230307A may have different origins: GRB 211211A could stem from an NS–WD merger, while GRB 230307A might arise from an NS–NS merger, with its prompt emission duration defined by mini-jet emission expected in the internal collision-induced magnetic reconnection and turbulence model \citep{zhang11,sun25,yi25a,yi25b,zhang25}.

Despite these efforts, several models face challenges. BH remnant–based models, whether NS–BH or NS–NS mergers, are difficult to explain the three distinct emission episodes (main burst: MB, extended emission: EE, and X-ray plateau exhibited by both GRBs 211211A and 230307A) with markedly different temporal and spectral behaviors using a single central engine, as emphasized by \cite{zhang25}. The NS–WD merger model of \cite{yang22} assumes that the NS merges into the center of the WD and triggers its collapse; however, this contradicts the conventional view in which the WD is completely tidally disrupted by the NS, as shown in hydrodynamic simulations \citep[e.g.,][]{met12,bob17,bob22,fer19,kal23}. 
On the other hand, the tidally disrupted NS–WD merger model of \cite{zhong23,zhong24} and \cite{wang24} cannot account for the r-process heavy elements identified by spectroscopic features in GRB 230307A \citep{levan24,gill23,gill25}, since nucleosynthesis in a disk composed of WD debris is expected to be proton-rich and incapable of producing robust r-process yields, according to hydrodynamic simulations of NS–WD mergers \citep[e.g.,][]{mar16,zen19,zen20,mor24,liu25}.

Nevertheless, the tidally disrupted NS–WD merger model of \cite{zhong23} successfully reproduces the three distinct emission episodes of GRB 211211A: the MB arises during the hyperaccretion phase onto the NS, the EE originates from the propeller phase, and the X-ray plateau is powered by the spin-down of a post-merger magnetar. 
Moreover, the pronounced trough between the MB and EE is naturally explained as the transition from accretion to propeller phases.
This model requires the pre-merger NS to already be a magnetar—a condition that has been demonstrated to be plausible in the precursor studies for both GRBs 211211A \citep{xiao24,suv22} and 230307A \citep{dich23}.
Given that nucleosynthesis in a disk composed of WD debris is incapable of producing robust r-process yields, \citet{zhong24} suggested the possibility of an alternative channel within the NS–WD merger framework to generate r-process heavy elements.

Motivated by recent work on r-process nucleosynthesis in magnetar giant flares (GFs), initially explored by \cite{cehula24}, confirmed by \cite{patel25a,patel25b}, applied to probe NS crusts by \cite{zhong26}, and further explored by \cite{qiu26} for focusing on MeV Gamma-ray lines of r-process nuclei, we propose that such flares may be naturally triggered during the accretion phase of a tidally disrupted NS–WD merger. If so, they could provide the baryonic ejecta with high entropy and rapid expansion necessary for r-process element production via the $\alpha$-rich freeze-out mechanism, 
thereby creating the observed kilonova signature.

\begin{figure*}[ht!]
	\includegraphics[width=1.\textwidth, angle=0]{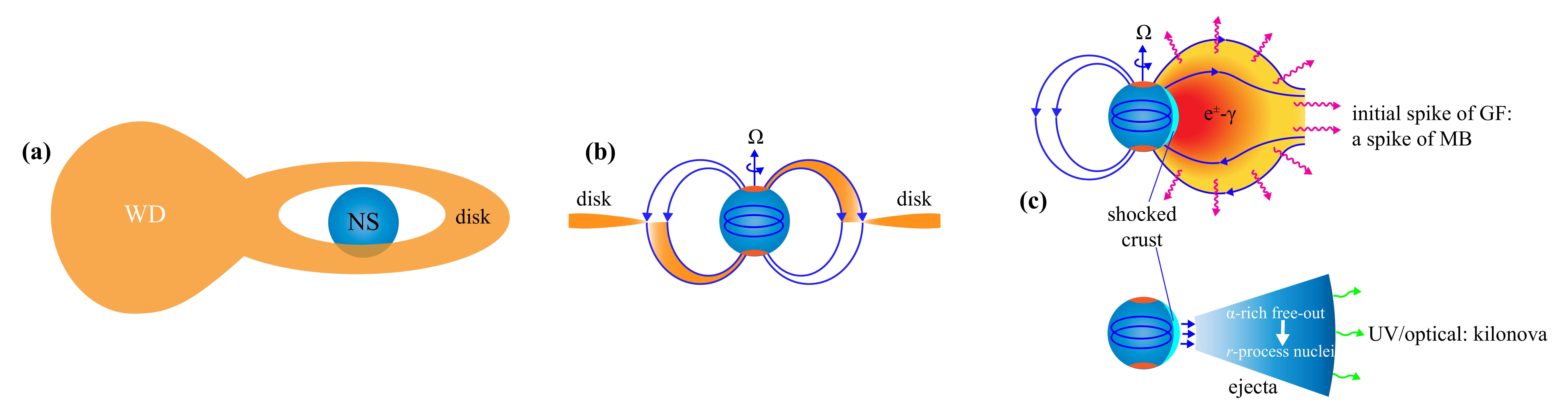}
	\caption{The unified picture for peculiar long gamma-ray bursts (LGRBs) such as GRBs~211211A and ~230307A from neutron star--white dwarf (NS--WD) mergers (see also Section \ref{sec:picture}).  
	Panel (a): A massive WD ($M_{\rm WD} \gtrsim 1\,M_\odot$) is tidally disrupted by a magnetar ($B_{\rm s} \gtrsim 10^{14}\,\mathrm{G}$), forming an accretion disk that feeds the NS at a rate $\gtrsim10^{-4}M_{\odot}~{\rm s}^{-1}$ lasting tens of seconds ($\gtrsim10^{-2}M_{\odot}~{\rm s}^{-1}$ lasting $\sim10$ s) described by hydrodynamic simulations in \cite{kal23}.  
	Panel (b): At an early stage, the NS is in the accretion phase ($r_{\rm m} \lesssim r_{\rm c}$), spinning up and amplifying its toroidal field from differential rotation via a dynamo mechanism \citep{zhong23}. This phase would give rise to the main burst (MB) of GRB prompt emission. At a late stage, the NS enters the propeller phase ($r_{\rm m} > r_{\rm c}$), expelling material to power the extended emission (EE) of prompt emission.  
	Panel (c): During the accretion phase, sustained mass inflow drives continuous amplification of the toroidal field, then the continuous rapid buildup of magnetic stresses in the NS crust triggers repeated starquakes, producing magnetar giant flares (GFs) preferentially in the closed-field-line region. Each initial spike in a GF corresponds to a high-pressure expanding electron/positron pair-photon ($e^\pm$-$\gamma$) fireball; this fireball drives a shock wave into the NS crust, leading to crustal ejection capable of \textit{r}-process nucleosynthesis via the $\alpha$-rich freeze-out mechanism \citep{cehula24}. Thousands of such events over $\sim 10-20$ s produce a forest of GF initial spikes that collectively form the MB, and yield $M_{\rm ej}\sim10^{-5}-10^{-3}\,M_\odot$ of ejecta that is sufficient to power the observed kilonova signature when further boosted by the spin-down of the post-merger magnetar. The ejecta mass can be significantly larger when accounting for the jet beaming effect. Meanwhile, the spin-down radiation also powers the X-ray plateau in GRB afterglow.
	\label{fig:unified} }
\end{figure*}

In this work, we therefore investigate how r-process heavy elements can be synthesized within the NS–WD merger framework of \cite{zhong23}, and present a unified picture capable of explaining the full set of observational features of peculiar LGRBs, particularly GRBs 211211A and 230307A, and applicable to future similar events.
The paper is organized as follows. Section~\ref{sec:picture} introduces the unified picture linking accretion dynamics to the production of the prompt emission, afterglow, and kilonova, building on the frameworks of \cite{zhong23} and \cite{cehula24}. Section~\ref{sec:explaining} outlines the comparative observational properties of GRBs 211211A and 230307A and reproduces these observations by our unified picture. A summary is provided in Section~\ref{sec:summary}.

\begin{table*}
	\centering
	\caption{Comparable observational properties of GRBs 211211A and 230307A}
	\label{tab:features}
	\renewcommand{\arraystretch}{1.4} 
	\begin{tabular}{l l l l}
		\hline\hline
		Feature & GRB 211211A$^a$ & GRB 230307A$^a$ & Reference$^b$ \\
		\hline
		\multicolumn{4}{l}{Prompt Emission} \\
		\hline
		Precursor & $T_{90} \sim 0.2~\mathrm{s}$, $E_{\rm p} \sim 54~\mathrm{keV}$ & $T_{90} \sim 0.4~\mathrm{s}$, $E_{\rm p} \sim 170~\mathrm{keV}$ & (1), (2) \\
		Main burst &
		\begin{tabular}[c]{@{}l@{}}$T_{90} \sim 13~\mathrm{s}$, $E_{\rm p} \sim 687~\mathrm{keV}$, $\tau_{\min} \sim 16~\mathrm{ms}$, \\ $E_{\gamma,\mathrm{iso}} \sim 5.30 \times 10^{51}~\mathrm{erg}$\end{tabular} &
		\begin{tabular}[c]{@{}l@{}}$T_{90} \sim 18~\mathrm{s}$, $E_{\rm p} \sim 800\text{--}900~\mathrm{keV}$, $\tau_{\min} \sim 9~\mathrm{ms}$, \\ ... \end{tabular} &
		(3), (2), (4) \\
		Extended emission & $T_{90} \sim 55~\mathrm{s}$, $E_{\rm p} \sim 82~\mathrm{keV}$ & $T_{90} \sim 27~\mathrm{s}$, $E_{\rm p} \sim 200\text{--}500~\mathrm{keV}$ & (3), (2) \\
		Trough & Yes & Yes & (5), (3), (2), (4) \\
		Whole burst &
		\begin{tabular}[c]{@{}l@{}}$L_{\gamma,\mathrm{p,iso}} \sim 1.94 \times 10^{51}~\mathrm{erg~s^{-1}}$, \\ $E_{\gamma,\mathrm{iso}} \sim 7.61 \times 10^{51}~\mathrm{erg}$\end{tabular} &
		\begin{tabular}[c]{@{}l@{}}$L_{\gamma,\mathrm{p,iso}} \sim 4.64 \times 10^{51}~\mathrm{erg~s^{-1}}$, \\ $E_{\gamma,\mathrm{iso}} \sim 3.18 \times 10^{52}~\mathrm{erg}$\end{tabular} &
		(3), (4) \\
		\hline
		\multicolumn{4}{l}{GRB Afterglow and Kilonova} \\
		\hline
		X-ray plateau & Yes & Yes & (6), (4) \\
		Kilonova & Candidate (no spectroscopic confirmation) & Confirmed (with spectroscopic features) & (5), (7), (8) \\
		\hline
	\end{tabular}
	\tablefoot{
		\tablefoottext{a}{$T_{90}$ is the duration; $E_{\rm p}$ is the peak energy; $\tau_{\min}$ is the minimum variability timescale; $L_{\gamma,\mathrm{p,iso}}$ is the peak isotropic luminosity; $E_{\gamma,\mathrm{iso}}$ is the isotropic radiation energy.} \\
		\tablefoottext{b}{(1) \citet{xiao24}; (2) \citet{dich23}; (3) \citet{yang22}; (4) \citet{sun25}; (5) \citet{ras22}; (6) \citet{tro22}; (7) \citet{levan24}; (8) \citet{gill23}.}
	}
\end{table*}

\section{The unified picture}\label{sec:picture} 
The unified picture is illustrated in Fig.~\ref{fig:unified} and described below. 

Panel (a): For an NS--WD merger involving a magnetar with surface magnetic field $B_{\rm s}\gtrsim10^{14}$ G and a massive WD with mass $M_{\rm WD}\gtrsim1M_{\odot}$ based on \cite{zhong23}, the WD is completely tidally disrupted by the NS, forming a disk. The NS subsequently accretes the disk material. The accretion profiles can be described by hydrodynamic simulations in \cite{kal23} for three types of disk prescriptions: high-entropy normal wind (HENW), constant entropy normal wind (CENW), and constant entropy efficient wind (CEEW). As shown in Fig.~3 of \cite{kal23}, for both the CENW and CEEW prescriptions, accretion rates $\gtrsim10^{-4}M_{\odot}~{\rm s}^{-1}$ persist for tens of seconds ($\gtrsim10^{-2}M_{\odot}~{\rm s}^{-1}$ persisting for $\sim10$ s). This is sufficient to account for the luminosity, isotropic radiation energy, and duration of the prompt emission of a peculiar LGRB such as GRB 211211A \citep[see Paragraph 3 of Section 2 in][]{zhong23}. 

Panel (b): According to the accretion profiles for the CENW and CEEW:
(1) At an early stage, the accretion rate is extremely high, leading to the magnetosphere radius $r_{\rm m}$ smaller than the corotation radius $r_{\rm c}$ ($r_{\rm m}\lesssim r_{\rm c}$).  
The NS enters the accretion phase, during which the disk material is funneled by the NS dipole field and accreted onto the stellar surface at the poles, causing the NS to spin up. 
(2) At a late stage, the accretion rate declines, $r_{\rm m}$ eventually exceeds $r_{\rm c}$ ($r_{\rm m} > r_{\rm c}$), the NS falls into the propeller phase, during which the material spins at a super-Keplerian rate to come into corotation with the NS and is thus expelled \citep{ill75}. 
During the accretion phase, differential rotation, potentially driven by instabilities such as the r-mode induced by accretion, 
amplifies the toroidal field via a dynamo mechanism, as previously applied by \cite{spr99} to X-ray binaries. 
This phase gives rise to the MB of prompt emission and the r-process ejecta, see Panel (c) below. 
Conversely, the propeller phase generates the EE of prompt emission through the propellered material, as studied in \cite{gom14} and \cite{gib17}.

Panel (c): In the accretion phase, if the NS has a surface field of $B_{\rm s}\sim10^{14}-10^{15}~{\rm G}$, a range roughly consistent with magnetic field estimates $\sim10^{15}~{\rm G}$ derived from precursor studies of GRBs 211211A and 230307A \citep{xiao24,suv22,dich23}, the toroidal field can rapidly amplify to $B_{\phi}\sim10^{15}-5\times10^{16}~{\rm G}$ \citep[see Fig.~1 in][]{zhong23}. Due to the toroidal field with the strongest amplification occurring on the closed-field-line region, the rapid buildup of magnetic stresses in the NS crust preferentially away from the symmetry axis rather than near it, would trigger a sudden starquake, leading to a large-scale reconnection of the stellar magnetic field \citep[see Section 3.1.1 in][]{thom01}. A magnetar GF may subsequently be created within the closed-field-line region, opening field lines during its prompt initial hard spike \citep[e.g.,][]{thom95,thom01,pern11,land15}. 
\cite{cehula24} proposed that during the GF initial spike, which is identified with an expanding electron/positron pair-photon ($e^{\pm}$-$\gamma$) fireball \citep{thom95}, the high-pressure fireball not only produces the prompt spike emission but also drives a shock wave into the NS crust. A fraction of the shocked neutron-rich crustal layers can become unbound, resulting in baryonic mass ejection. Although the ejecta has a relatively high electron fraction $Y_e\gtrsim0.4$, its high entropy and rapid expansion enable an $\alpha$-rich freeze-out process, facilitating the synthesis of r-process heavy elements. The entire sequence from fireball launch to ejecta escaping occurs on a timescale of $\lesssim1$ ms, estimated from \cite{cehula24}, who showed that the unbound ejecta mass ultimately saturates by $\sim600\mu{\rm s}$ for the fiducial model presented in their Section 4.1. Each fireball thus corresponds to one GF initial spike and one ejecta parcel with mass $m_{\rm ej}\sim10^{-8}-10^{-6}M_{\odot}$, see Table 1 of \cite{cehula24}.

Given that the accretion phase lasts $\sim10-20$ s, matching the MB duration of GRB 211211A or GRB 230307A, thousands of such fireballs can be produced in succession. Their collective emission composed of a superposition--a ``forest'' of GF initial spikes forms the observed MB\footnote{Notably, this picture does not require the toroidal field to exceed the buoyancy limit $B_{\rm b}\sim10^{17}$ G \citep{klu98}, a condition normally required for magnetic toroids to penetrate through the stellar surface and reconnect to form magnetic bubbles. The magnetic bubble model was previously invoked by \cite{zhong23} to explain the MB.}. Moreover, the cumulative mass of thousands of ejecta parcels reaches $M_{\rm ej}\sim10^{-5}-10^{-3}M_{\odot}$, sufficient to power a kilonova when further enhanced by the spin-down of the post-merger magnetar. Notably, the jet half-opening angles are estimated to be $\theta_{\rm j} \sim 0.01-0.05$~rad for GRB~211211A \citep{ras22,yang22,tro22,mei22,zhong23} and $\sim 0.05-0.08$~rad for GRB~230307A \citep{yang24,sun25,zhong24}, corresponding to a monopolar beaming factor $f_{\rm b}\simeq 0.5\times\frac{1}{2}\theta_{\rm j}^2\sim10^{-5}-10^{-3}$. Consequently, the average half-opening angle of the GF initial spikes constituting the MB should align with the corresponding jet angle $\theta_{\rm j}$. Even if the majority of GF initial spikes are preferentially generated in the closed-field-line region or even narrowly confined to the equator as assumed in our unified picture, those originating from the far side or lateral edges of the star may remain outside our line of sight (LOS). Therefore, the inferred number of consecutive spikes derived from the MB duration for these two GRBs in Section \ref{subsec:prompt} represents merely a lower limit. In this case, the inferred cumulative ejecta mass of $M_{\rm ej} \sim 10^{-5}-10^{-3}~M_\odot$ should also be regarded as a lower limit.

Additionally, the spin-down radiation also powers the X-ray plateau in GRB afterglow, as demonstrated in \cite{zhong23,zhong24}, \cite{yang22}, and \cite{sun25}.

\section{Explaining peculiar long gamma-ray bursts}\label{sec:explaining}
Peculiar LGRBs such as GRBs 211211A and 230307A share several key observational features listed in Table \ref{tab:features} and described below. 
\begin{enumerate}
	\item The prompt emission is long-duration and consists of three distinct components: a soft-spectrum precursor \citep{xiao24,dich23}, a hard-spectrum MB \citep{yang22,dich23}, and a soft-spectrum EE\footnote{The temporally extended and spectrally soft tail post the main long-duration spectrally hard burst can be regarded as an EE for GRB 230307A \citep{dich23}.} \citep{yang22,dich23}. The MB and EE are separated by a pronounced trough. Moreover, their whole bursts exhibit comparable peak isotropic luminosities $L_{\gamma,\mathrm{p,iso}}$ and isotropic radiation energies $E_{\gamma,\mathrm{iso}}$.  
	Based on Section 2 of \cite{zhong23}, the prompt emission containing the MB and the EE can be directly related to the accretion profile in the CEEW or CENW prescription of \cite{kal23}. Specifically, for GRB 211211A, the observed $L_{\gamma,\rm p,iso}$, $E_{\gamma,\rm iso}$, and $T_{90}$ duration of its prompt emission can be all reproduced by the accretion profile of an $M_{\rm NS}:M_{\rm WD}=1.40M_{\odot}:1.00M_{\odot}$ binary. GRB~230307A displays comparable luminosity, energy, and duration, and is likewise well accounted for by the same binary configuration.
	
	\item An X-ray plateau is observed in GRB afterglow, which is commonly interpreted as a signature of a post-merger magnetar \citep{yang22,tro22,zhong23,zhong24,sun25}. 
	
	\item A kilonova signature is present in both events, albeit with initially differing interpretations. GRB~211211A lacks spectroscopic confirmation of r-process heavy elements. Consequently, \cite{zhong23} previously proposed its kilonova-like emission may arise from a magnetar-boosted rapidly evolving transient powered by $^{56}\mathrm{Ni}$ decay. In contrast, GRB~230307A exhibits clear spectroscopic signatures of r-process elements \citep{levan24,gill23,gill25}, confirming a genuine r-process-powered kilonova. 
	Given the numerous shared properties between the two events, we adopt the unified model illustrated in Fig.~\ref{fig:unified}, in which r-process ejecta are created during magnetar GFs. Under this framework, we interpret the kilonova signal in GRB~211211A as a realistic r-process kilonova.
\end{enumerate}

\subsection{Main burst: a forest of giant flare initial spikes}\label{subsec:prompt}
Within the unified picture in Fig. \ref{fig:unified}, the MB is interpreted as a forest of initial spikes from repeated GFs. The observed minimum variability timescales, $\tau_{\min} \sim 16\,\mathrm{ms}$ for GRB~211211A and $\tau_{\min} \sim 9\,\mathrm{ms}$ for GRB~230307A, imply that the MB durations ($T_{90} \sim 13\,\mathrm{s}$ and $\sim18\,\mathrm{s}$, respectively) consist of $\sim800$ and $\sim2000$ consecutive spikes. The minimum variability timescale of $\sim 10\,\mathrm{ms}$, interpreted as the duration of a single spike in the MB, 
is comparable to the durations of initial spikes observed in magnetar GFs from the Milky Way and nearby galaxies, such as the SGR~1806–20 GF \citep{hur05,palm05,fred07a}, GRB~051103 \citep{ofek06,fred07b}, GRB~070201 \citep{ofek08,maz08}, and GRB~200415A \citep{svin21,robe21,yang20,zhang20}, when cutting their long decay phases, see Table~4 of \citet{yang20}\footnote{Also can see \cite{zhang20} for a larger sample.}. Furthermore, based on the observed decay durations and distances in their table, GF initial spikes at greater distances generally exhibit shorter decay phases. This indicates that the intrinsically long decay phase of a distant initial spike should suffer from severe background burial. Similarly, in the MB of GRB~ 211211A or GRB~230307A, the intrinsically long decay phase of a pre-spike should be obscured by continuous post-spikes rather than background. Therefore, equating the main burst minimum variability timescale with the duration of a single GF initial spike is reasonable.

For GRB~211211A, the isotropic energy of the MB comprising $\sim 800$ spikes is $E_{\gamma,\mathrm{iso}} \sim 5.30 \times 10^{51}\,\mathrm{erg}$, comparable to the total energy of the whole burst. This yields an average isotropic energy per spike of $E_{\gamma,\mathrm{iso,ave}}^{\rm spike} \sim 6.6 \times 10^{48}\,\mathrm{erg}$. Given a spike duration of $\tau_{\min} \sim 16\,\mathrm{ms}$, the corresponding average isotropic luminosity is $L_{\gamma,\mathrm{iso,ave}}^{\rm spike} \sim 4.1 \times 10^{50}\,\mathrm{erg\,s^{-1}}$. Similarly, for GRB~230307A, each spike carries $E_{\gamma,\mathrm{iso,ave}}^{\rm spike} \sim 1.6 \times 10^{49}\,\mathrm{erg}$ and $L_{\gamma,\mathrm{iso,ave}}^{\rm spike} \sim 1.8 \times 10^{51}\,\mathrm{erg\,s^{-1}}$. 
Notably, the isotropic energies and luminosities of individual MB spikes in both GRBs exceed those of known Galactic and nearby extragalactic GFs, which result from evolved magnetars, by $\sim2$–4 orders of magnitude (see also Table~4 in \citealt{yang20}), suggesting a significantly stronger magnetic field than that inferred for such evolved magnetars. 
If the spike energy originates solely from the magnetic energy of the accretion-amplified toroidal field, then using the relation $E_{\gamma,\mathrm{iso,ave}}^{\rm spike} \sim \frac{B_\phi^2}{8\pi} \Delta R^3$ with fireball scale $\Delta R\sim R$ where an NS radius $R \approx 12\,\mathrm{km}$, one can infer a toroidal field strength of a few $\times 10^{16}\,\mathrm{G}$. This is consistent with the range $B_\phi \sim 10^{15}-5 \times 10^{16}\,\mathrm{G}$ in Fig.~1 of \citet{zhong23}.

\subsection{Kilonova: r-process during a forest of giant flare initial spikes}\label{subsec:kilonova}
According to the unified picture in Fig.~\ref{fig:unified}, the kilonova signatures of both GRBs~211211A and 230307A are interpreted as being powered by radioactive decay of \textit{r}-process heavy elements synthesized in unbound baryonic ejecta. This ejecta consists of thousands of discrete parcels launched during a forest of initial spikes from GFs that collectively form the MB. 
As shown in Table~1 of \citet{cehula24}, each spike is associated with an ejecta parcel of mass $m_{\rm ej} \sim 10^{-8}-10^{-6}\,M_{\odot}$ and average velocity $v \sim 0.1-0.4\,c$, if a volume-averaged magnetic field $B_{\rm ave} \sim B_{\phi} \sim 10^{15}-5\times10^{16}\,\mathrm{G}$. For thousands of such parcels, the cumulative ejecta mass reaches $M_{\rm ej} \sim 10^{-5}-10^{-3}\,M_{\odot}$.

Although the density and velocity structure of the total ejecta is expected to be highly complex, even for a single parcel as exhibited in \citet{cehula24} and \citet{patel25a,patel25b}, we adopt a simplified model for the kilonova emission. Assuming centrally located energy deposition and homologous expansion, the bolometric luminosity of the kilonova boosted by magnetar spin-down can be approximated using a semi-analytical formalism originally developed for supernovae \citep[SNe; e.g.,][]{arnett82,kasen10,woo10,chatz12,lilong20}, but with $^{56}\mathrm{Ni}$ replaced by \textit{r}-process material \citep[e.g.,][]{yu13,met14b}:
\begin{eqnarray}
	\begin{aligned}
		L(t)=& \frac{2}{\tau_{\rm d}} e^{-\frac{t^{2}}{\tau_{\rm d}^2}}  \\
		\times& \int_{0}^{t} e^{\frac{t^{\prime 2}}{\tau_{\rm d}^2}} \left(\frac{t^{\prime}}{\tau_{\rm d}}\right) P(t^{\prime}) d t^{\prime}\ \mathrm{erg~s}^{-1},
	\end{aligned}
	\label{eq:Lt}
\end{eqnarray}
where $\tau_{\rm d}$ is the photon diffusion timescale, given by
\begin{equation}
	\tau_{\rm d} = \left( \frac{2 \kappa M_{\rm ej}}{\beta v c} \right)^{1/2},
	\label{eq:tau_m}
\end{equation}
with $\kappa \sim 3-20~\mathrm{cm^2~g^{-1}}$ denoting the opacity, which varies depending on the composition of \textit{r}-process nuclei from second-peak/light elements to third-peak species \citep{patel25a}. Here $v$ is the characteristic expansion velocity and $\beta \approx 13.7$ is a dimensionless constant related to the ejecta's geometric density profile \citep{arnett82}.

The total power input $P(t)$ comprises two components
\begin{equation}
	P(t) = P_{\rm rp}(t) + P_{\rm mag}(t),
	\label{eq:Pt}
\end{equation}
where the first term represents the radioactive power from \textit{r}-process material \citep{patel25b}
\begin{equation}
	P_{\rm rp}(t) = M_{\rm rp} f_{\rm th} \dot{q}_r(t),
	\label{eq:P_rp}
\end{equation}
with $M_{\rm rp} \sim 0.3-0.8\,M_{\rm ej}$ being the mass of \textit{r}-process nuclei, $f_{\rm th} \sim0.6-0.7$ the thermalization efficiency, and
\begin{equation}
	\dot{q}_r(t) \approx 5 \times 10^{12} \left( \frac{t}{10^3~\mathrm{s}} \right)^{-\alpha} \mathrm{erg~s^{-1}~g^{-1}}
	\label{eq:dot_qr}
\end{equation}
the specific radioactive energy generation rate. The exponent $\alpha = 1.2$ is adopted based on nucleosynthesis calculations in \citet{patel25a,patel25b}. In our joint modeling below, we fix $M_{\rm rp} = 0.5\,M_{\rm ej}$ and $f_{\rm th} = 0.7$.

The second term accounts for additional energy injection from magnetar spin-down \citep{dai98a,dai98b,zhang01}
\begin{equation}
	P_{\rm mag}(t) = \frac{E_{\rm mag}}{\tau_{\rm mag}} \left( 1 + \frac{t}{\tau_{\rm mag}} \right)^{-2},
	\label{eq:P_ns}
\end{equation}
where $E_{\rm mag} = \frac{1}{2} I \Omega_0^2$ is the initial rotational energy of the magnetar, linked to its spin period $P_0 = 2\pi / \Omega_0$. The characteristic spin-down timescale is $\tau_{\rm mag} = 3 I c^3 / (B_{\rm s,post}^2 R^6 \Omega_0^2)$, with the moment of inertia approximated as $I = \frac{2}{5} M_{\rm NS} R^2$. Here $B_{\rm s,post}$ is the surface field of the post-merger magnetar.

To model the multi-wavelength light curves of kilonova emission, we compute the observed flux density $F_\nu(t)$ in units of $\mathrm{erg~s^{-1}~cm^{-2}~Hz^{-1}}$ from the photospheric temperature evolution. The temperature is derived from the bolometric luminosity via
\begin{equation}
	T(t) = \left[ \frac{L(t)}{4\pi \sigma_{\rm SB} v^2 t^2} \right]^{1/4},
	\label{eq:T}
\end{equation}
and the flux density follows from blackbody radiation
\begin{equation}
	F_{\nu}(t)=\frac{2 \pi h \nu^{3}}{c^{2}} \frac{1}{e^{h \nu / k T}-1}\frac{v^2t^2}{D_{\rm L}^2},
	\label{eq:F_v}
\end{equation}
where $D_{\rm L}$ is the luminosity distance to the source and $\sigma_{\rm SB}$ is the Stefan–Boltzmann constant. Finally, the monochromatic AB magnitude is obtained as
\begin{equation}
	M_\nu(t) = -2.5 \log_{10} \left[ \frac{F_{\nu}(t)}{3631~\mathrm{Jy}} \right].
\end{equation}

\begin{figure*}[ht!]
	\includegraphics[width=0.5\textwidth, angle=0]{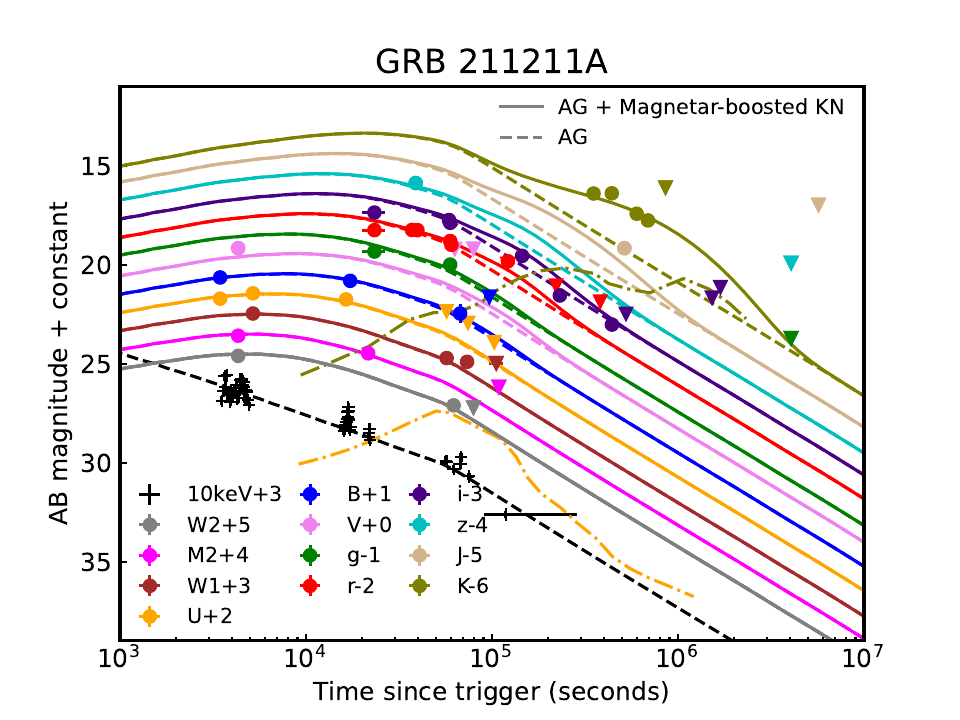}
	\includegraphics[width=0.5\textwidth, angle=0]{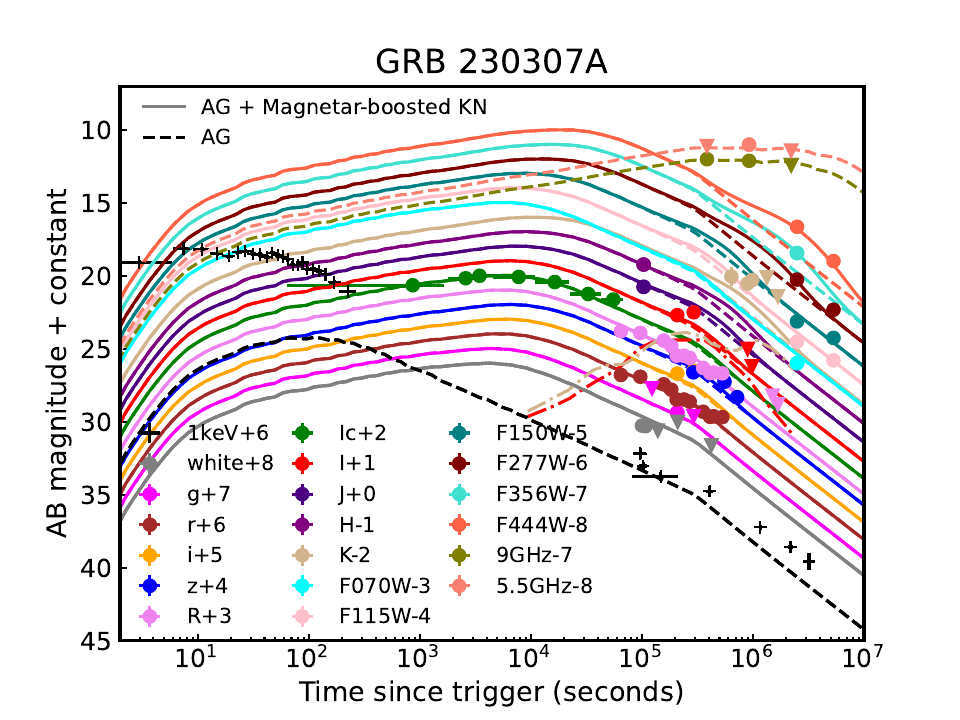}
	\caption{Joint modeling of the multi-wavelength light curves of the GRB afterglow and kilonova signature for GRB 211211A (left panel) and GRB 230307A (right panel), combining the standard afterglow (AG) model with the magnetar-boosted kilonova (KN) model. Dashed fitting lines show the contribution from the standard afterglow model alone, while solid lines represent the full model that includes both the afterglow and the magnetar-boosted kilonova components. Inverted triangles denote upper limits. The dashed-dotted lines (left panel: orange for U band, olive for K band; right panel: red for I band, tan for K band) represent the subluminous optical transient from $^{56}{\rm Ni}$ decay in the disk wind for the CEEW scenario , adapted from the third panel of Figure 12 in \cite{kal23} for $M_{\rm NS}:M_{\rm WD}=1.10M_{\odot}:1.25M_{\odot}$ binary. 
		\label{fig:kilonova}}
\end{figure*}

\begin{table}
	\centering
	\caption{Best-fit results for jointly modeling multi-wavelength GRB afterglow and kilonova}
	\label{tab:parameters}
	\renewcommand{\arraystretch}{1.4} 
	\begin{tabular}{l l l}
		\hline\hline
		Parameter & GRB 211211A & GRB 230307A \\
		\hline
		\multicolumn{3}{l}{GRB Afterglow} \\
		\hline
		$\theta_{\rm j}$ (rad) & $0.028^{+0.003}_{-0.002}$ & $0.072^{+0.006}_{-0.006}$ \\
		$\log_{10}(E_{\rm K,iso}/{\rm erg})$ & $50.78^{+0.07}_{-0.09}$ & $50.72^{+0.12}_{-0.11}$ \\
		$\log_{10}(\Gamma_0)$ & $2.36^{+0.11}_{-0.11}$ & $2.77^{+0.19}_{-0.17}$ \\
		$\log_{10}(n/{\rm cm}^{-3})$ & $-5.71^{+0.21}_{-0.16}$ & $-4.58^{+0.26}_{-0.21}$ \\
		$\log_{10}(\epsilon_B)$ & $-0.59^{+0.05}_{-0.09}$ & $-0.70^{+0.15}_{-0.34}$ \\
		$\log_{10}(\epsilon_e)$ & $-0.51^{+0.01}_{-0.01}$ & $-0.63^{+0.06}_{-0.03}$ \\
		$p$ & $2.83^{+0.02}_{-0.02}$ & $2.82^{+0.02}_{-0.02}$ \\
		\hline
		\multicolumn{3}{l}{Magnetar-boosted Kilonova} \\
		\hline
		$P_0$ (ms) & $141.82^{+19.56}_{-26.42}$ & $255.42^{+113.08}_{-58.30}$ \\
		$B_{\rm s,post}$ ($10^{14}~{\rm G}$) & $6.66^{+1.87}_{-2.03}$ & $4.64^{+2.83}_{-1.83}$ \\
		$v$ ($c$) & $0.34^{+0.04}_{-0.04}$ & $0.11^{+0.01}_{-0.01}$ \\
		$\log_{10}(M_{\rm ej}/M_{\odot})$ & $-1.91^{+0.06}_{-0.09}$ & $-1.91^{+0.08}_{-0.03}$ \\
		$\kappa$ (cm$^2$ g$^{-1}$) & $0.88^{+0.14}_{-0.05}$ & $4.28^{+1.36}_{-1.26}$ \\
		\hline
	\end{tabular}
\end{table}

\subsection{Jointly modeling GRB afterglow and kilonova}\label{subsec:joint}
For GRB 211211A, the observed optical and near-infrared data displayed in Fig.~\ref{fig:kilonova} are taken from \cite{ras22}, while the Swift UVOT/XRT light curves are adopted from \cite{rom05} and \cite{evans09}. For GRB 230307A, the early X-ray data (1keV) before 300 s are scaled from the Lobster Eye Imager for Astronomy \citep[LEIA;][]{sun25}, the late are collected from the Swift XRT, XMM-Newton, and Chandra \citep{yang24}; while the optical and radio data are drawn from \cite{levan24}. 
Since the datasets should encompass both the GRB afterglow and the kilonova emission, 
we perform a joint fit to the multi-wavelength afterglow and kilonova data by combining the standard afterglow model \citep{sari98,huang99} with the magnetar-boosted kilonova model described in Section~\ref{subsec:kilonova}.  
The free parameters of the joint modeling include: the jet half-opening angle $\theta_{\rm j}$, the isotropic kinetic energy $E_{\rm K,iso}$, the initial Lorentz factor $\Gamma_0$, the fractions of shock energy channeled into the magnetic field $\epsilon_B$ and electrons $\epsilon_e$, the circumburst medium density $n$, the power-law index $p$ of the electron energy distribution, the initial spin period $P_0$ and surface magnetic field $B_{\rm s,post}$ of the post-merger magnetar, as well as the ejecta mass $M_{\rm ej}$, expansion velocity $v$, and opacity $\kappa$.

Note that here we exclude the contribution from disk wind ejecta in our modelings. On one hand, as mentioned in Section 5.1 of \cite{kal23}, because the inner wind material quickly leads the wind shock and becomes optically thin due to high velocity, most of the gamma rays in the $^{56}{\rm Ni}$ decay in the disk wind ejecta may escape directly rather than being thermalized, producing gamma-ray emission with a peak luminosity in the range of $10^{39}-10^{42}~{\rm erg~s^{-1}}$. However, this gamma-ray emission cannot be observed by the current gamma-ray instruments due to its large distance for GRB~211211A or GRB~230307A. On the other hand, according to Table~1 in both \cite{zhong23} and \cite{wang24}, $\sim10^{-3}~M_{\odot}$ of $^{56}{\rm Ni}$ can reproduce the observed kilonova-like emission in GRB~211211A and GRB~230307A. Therefore, as long as the actual thermalized $^{56}{\rm Ni}$ mass $<10^{-3}~M_{\odot}$, one can exclude the contribution from the $^{56}{\rm Ni}$-bearing disk wind. This is supported by the results in Figure~\ref{fig:kilonova}, where the dashed-dotted lines (left: orange/U and olive/K; right: red/I and tan/K), which represent subluminous optical transients from the thermalized $^{56}\mathrm{Ni}$ decay in the CEEW disk wind adapted from the third panel of Figure~12 of \cite{kal23} for $M_{\rm NS}:M_{\rm WD}=1.10M_{\odot}:1.25M_{\odot}$ binary, lie below the solid lines with the same bands. 

To identify the best-fit parameter values, we employ the Markov chain Monte Carlo (MCMC) algorithm, sampling within reasonable prior ranges. The resulting best-fit light curves are shown in Fig.~\ref{fig:kilonova}, while the inferred parameter values are listed in Table~\ref{tab:parameters}. 
Our main findings are summarized as follows:
\begin{enumerate}
	\item The best fit successfully reproduces the multi-wavelength light curves of both the GRB afterglow and the kilonova for both GRBs (see Fig.~\ref{fig:kilonova}). The derived ejecta properties such as mass $M_{\rm ej}$, velocity $v$, and opacity $\kappa$ fall within the ranges predicted by hydrodynamical simulations \citep{cehula24} and nuclear reaction network calculations \citep{patel25a}. The standard afterglow parameters are consistent with typical values observed in SGRBs. Moreover, the inferred magnetar parameters align broadly with those reported in Table 1 of \cite{zhong23}. Notably, the post-merger magnetar’s surface field, $B_{\rm s,post} \sim 4-7 \times 10^{14}\,\mathrm{G}$ for both GRBs, lies within the expected $B_{\rm s}\sim10^{14}-10^{15}\,\mathrm{G}$ range during merger, in line with the scenario in which accretion predominantly amplifies the toroidal component of the field.
	
	\item For both GRBs~211211A and 230307A, the magnetar parameters initial spin period $P_0$ and post-merger surface field $B_{\rm s,post}$ inferred from kilonova modeling differ from those derived from the X-ray plateau analysis in \citet{zhong23,zhong24}. As suggested by \citet{zhong23}, this discrepancy may be reconciled if the magnetar’s spin-down radiation exhibits a structured angular distribution: the dominant component is collimated along the GRB jet, powering the X-ray plateau, while a subdominant quasi-isotropic component injects energy into the surrounding ejecta, thereby boosting the kilonova emission.
	Moreover, in the unified picture illustrated in Fig.~\ref{fig:unified}, the forest of GF initial spikes originates from the closed-field-line region of the magnetar and form the GRB jet. Consequently, the bulk of the spin-down radiation beamed along the jet is also expected to emerge from this same region, consistent with the angular profile of magnetic dipole radiation from an orthogonal rotator.
	
	\item The high opacity $\kappa\sim4.28~{\rm cm^2~g^{-1}}$ for GRB 230307A, corresponding to ejecta rich in second/third peak r-process nuclei, is physically justified: (1)  \cite{patel25a} confirmed, using nuclear reaction network calculations, that the GF ejecta synthesize moderate yields of third-peak r-process nuclei and more substantial yields of second-peak/light r-nuclei, as shown in total nucleosynthetic yields in the top panel of Figure 3 of \cite{patel25a}. (2) For the kilonova AT 2017gfo \citep[e.g.,][]{kasen17}, its early emission originates from lanthanide-poor material, while its late emission is lanthanide-dominated. Similarly, for GRB 230307A, its early light r-process emission is hidden by the GRB afterglow, and the observed late-time kilonova flux arises from second/third peak material, which outshines the fading GRB afterglow. This is consistent with the tellurium identification reported in \cite{levan24} and the lanthanide spectral features found in \cite{gill23}.
	
	\item From the modeling results in Table 2, both GRBs favor $M_{\rm ej}\sim 10^{-2}~M_\odot$. This large ejecta mass can be explained by the jet beaming effect, albeit requiring at least ten times more misaligned GF events than the observable along our LOS, according to the second to last paragraph in Section \ref{sec:picture}. 
\end{enumerate}

\section{Summary}\label{sec:summary} 
Peculiar LGRBs, exemplified by GRBs~211211A and 230307A, exhibit SGRB-like characteristics yet possess long durations, challenging the traditional dichotomy in GRB classification. These events display a multi-component prompt emission comprising a precursor, a MB, and an EE, accompanied by an X-ray plateau in the afterglow and an associated kilonova signature. Their origin remains highly debated.
In this work, we have proposed a unified picture for these peculiar LGRBs based on tidally disrupted NS--WD mergers involving a pre-merger magnetar with $B_{\rm s} \gtrsim 10^{14}\,\mathrm{G}$ and a massive WD with $M_{\rm WD} \gtrsim 1\,M_{\odot}$ \citep{zhong23}. The plausibility of such a pre-merger magnetar has been supported by precursor studies of both GRBs~211211A and 230307A \citep{xiao24,dich23}.

This picture naturally accounts for the full suite of observed features including the prompt emission, X-ray plateau, and kilonova signature in both events. In the picture, the massive WD is fully tidally disrupted by the NS, forming a constant-entropy accretion disk that drives hyperaccretion onto the NS at a rate $\gtrsim 10^{-4}\,M_\odot\,\mathrm{s}^{-1}$ for tens of seconds \citep[$\gtrsim10^{-2}M_{\odot}~{\rm s}^{-1}$ lasting $\sim10$ s;][]{kal23}. During the early accretion phase, magnetically channeled material spins up the NS and amplifies its toroidal magnetic field to $B_{\phi} \sim 10^{15}$--$5 \times 10^{16}\,\mathrm{G}$ via a dynamo mechanism \citep{zhong23}. The continuous buildup of magnetic stresses triggers repeated starquakes, producing magnetar GFs. Each initial spike of a GF is associated with an expanding $e^\pm$-$\gamma$ fireball. This fireball shocks the NS crust, leading to $m_{\rm ej} \sim 10^{-8}$--$10^{-6}\,M_{\odot}$ of crustal ejecta parcel that synthesizes \textit{r}-process heavy elements through the $\alpha$-rich freeze-out mechanism \citep{cehula24}. Thousands of such successive GF spikes toward our LOS collectively form the MB. As the accretion rate declines, the NS transitions into the propeller phase, giving rise to the EE. The trough observed between the MB and EE thus arises naturally from this phase transition. Over the MB duration, the cumulative mass of thousands of ejecta parcels reaches $M_{\rm ej} \sim 10^{-5}$--$10^{-3}\,M_{\odot}$, sufficient to power the observed kilonova when further enhanced by the spin-down from the post-merger magnetar. The ejecta mass can be significantly larger when accounting for the jet beaming effect. Meanwhile, the spin-down radiation also powers the X-ray plateau. 

Within this picture, the MBs of GRBs~211211A and 230307A originate from a forest of initial spikes from repeated magnetar GFs. Their minimum variability timescales, $\tau_{\min} \sim 16\,\mathrm{ms}$ and $\sim 9\,\mathrm{ms}$, imply $\sim800$ and $\sim2000$ spikes over MB durations of $T_{90} \sim 13\,\mathrm{s}$ and $\sim18\,\mathrm{s}$, respectively, consistent with the spike widths seen in Galactic and extragalactic GFs once long decay tails are excluded. The average isotropic energy (luminosity) per spike is $E_{\gamma,\mathrm{iso,ave}}^{\mathrm{spike}} \sim 6.6 \times 10^{48}\,\mathrm{erg}$ ($L_{\gamma,\mathrm{iso,ave}}^{\mathrm{spike}} \sim 4.1 \times 10^{50}\,\mathrm{erg~s^{-1}}$) for GRB~211211A and $\sim 1.6 \times 10^{49}\,\mathrm{erg}$ ($\sim 1.8 \times 10^{51}\,\mathrm{erg~s^{-1}}$) for GRB~230307A, exceeding those of known GFs from evolved magnetars by 2--4 orders of magnitude. This suggests that the magnetic field is significantly stronger than the fields inferred for such evolved magnetars. If the spike energy originates solely from the magnetic energy of the accretion-amplified toroidal field, one can infer a toroidal field strength of a few $\times 10^{16}\,\mathrm{G}$. 

Accordingly, the kilonovae associated with GRBs~211211A and 230307A are considered to be powered by the radioactive decay of \textit{r}-process nuclei synthesized in thousands of discrete ejecta parcels. We jointly model the multi-wavelength light curves of the afterglow and kilonova by combining the standard external shock model with a semi-analytical magnetar-boosted kilonova formalism using the MCMC method. The best fit successfully reproduces the observed light curves for both GRBs. The inferred ejecta mass, velocity, and opacity align with hydrodynamic simulations and nucleosynthesis calculations, while the derived afterglow parameters are consistent with those of typical SGRBs. Moreover, the post-merger magnetar's surface field, $B_{\rm s,post} \sim 4-7 \times 10^{14}\,\mathrm{G}$ for both events, falls within the expected range of $B_{\rm s}\sim10^{14}$--$10^{15}\,\mathrm{G}$ during merger, in line with the scenario that accretion predominantly amplifies the toroidal component of the magnetic field.

Several caveats stemming from our simplifying assumptions warrant mention.
First, as discussed in \citet{cehula24}, their model assumed a cold (i.e., non-accreting) radial structure for the NS crust prior to a GF. In our scenario, however, the NS is actively accreting, and its surface, particularly near the magnetic poles where infalling material passes through a shock before settling, should be hot \citep[see Appendix in][]{piro11}. It remains uncertain whether this hot layer can spread sufficiently to cover the closed-field-line region within the $\sim 10\,\mathrm{s}$ accretion interval. If not, the cold-crust assumption may still hold, given that the fireballs responsible for GF initial spikes are generally launched from the closed-field-line region.
Second, \citet{cehula24} cautioned that a cold crust may not be a good approximation when multiple GFs occur in rapid succession, as could happen under sustained accretion, since repeated ejections might outpace the crust’s ability to re-establish $\beta$-equilibrium. This effect is not incorporated in our current treatment and will be addressed in forthcoming work by \citet{cehula24}.

\begin{acknowledgements}
This work is supported by the National Natural Science Foundation of China (No. 12503048), the Guangxi Natural Science Foundation (No. 2026GXNSFBA00640119),
the Starting Foundation of Guangxi University of Science and Technology (No. 24Z17), and the Guangxi Qingmiao Research Program. L.L. is supported by the National Natural Science Foundation of China (Nos. 12563009, 12303050) and the Jiangxi Provincial Natural Science Foundation (No. 20252BAC200593). L.Z. is supported by the National Natural Science Foundation of China (No. 12403056). Y.Z.M. is supported by the National Natural Science Foundation of China (No. 12403045), the Natural Science Foundation of Guangxi (No. 2025GXNSFBA069091), and the Starting Foundation of Guangxi University of Science and Technology (No. 24Z01).
We also acknowledge our use of public data from the Swift, Fermi, Einstein Probe, XMM-Newton, and Chandra data archive.
\end{acknowledgements}

\bibliographystyle{aa}
\bibliography{refs}

@ARTICLE{zhong26,
	author = {{Zhong}, Jiahang and {Chen}, Qiu-Hong and {Kang}, Yacheng and {Li}, Hong-Bo and {Zhang}, Jinghao and {Chen}, Meng-Hua and {Shao}, Lijing},
	title = "{Novae breves from magnetar giant flares: Potential probes of neutron star crusts}",
	journal = {\aap},
	year = 2026,
	month = may,
	volume = {709},
	eid = {A195},
	pages = {A195},
	doi = {10.1051/0004-6361/202659203},
	archivePrefix = {arXiv},
	eprint = {2603.10500},
	primaryClass = {astro-ph.HE},
	adsurl = {https://ui.adsabs.harvard.edu/abs/2026A&A...709A.195Z}
}

@ARTICLE{gill25,
	author = {{Gillanders}, J.~H. and {Smartt}, S.~J.},
	title = "{Analysis of the JWST spectra of the kilonova AT 2023vfi accompanying GRB 230307A}",
	journal = {\mnras},
	year = 2025,
	month = apr,
	volume = {538},
	number = {3},
	pages = {1663-1689},
	doi = {10.1093/mnras/staf287},
	archivePrefix = {arXiv},
	eprint = {2408.11093},
	primaryClass = {astro-ph.HE},
	adsurl = {https://ui.adsabs.harvard.edu/abs/2025MNRAS.538.1663G}
}

@ARTICLE{qiu26,
	author = {{Qiumu}, Wu-Zimo and {Chen}, Meng-Hua and {Chen}, Qiu-Hong and {Xie}, Fei and {L{\"u}}, Hou-Jun and {Wang}, Xiang-Gao and {Liang}, En-Wei},
	title = "{MeV Gamma-Ray Lines from Radioactive Nuclei in Magnetar Giant Flares}",
	journal = {arXiv e-prints},
	year = 2026,
	month = jun,
	eid = {arXiv:2606.17997},
	pages = {arXiv:2606.17997},
	doi = {10.48550/arXiv.2606.17997},
	archivePrefix = {arXiv},
	eprint = {2606.17997},
	primaryClass = {astro-ph.HE},
	adsurl = {https://ui.adsabs.harvard.edu/abs/2026arXiv260617997Q}
}

@ARTICLE{land15,
	author = {{Lander}, S.~K. and {Andersson}, N. and {Antonopoulou}, D. and {Watts}, A.~L.},
	title = "{Magnetically driven crustquakes in neutron stars}",
	journal = {\mnras},
	year = 2015,
	month = may,
	volume = {449},
	number = {2},
	pages = {2047-2058},
	doi = {10.1093/mnras/stv432},
	archivePrefix = {arXiv},
	eprint = {1412.5852},
	primaryClass = {astro-ph.HE},
	adsurl = {https://ui.adsabs.harvard.edu/abs/2015MNRAS.449.2047L}
}

@ARTICLE{pern11,
	author = {{Perna}, Rosalba and {Pons}, Jose A.},
	title = "{A Unified Model of the Magnetar and Radio Pulsar Bursting Phenomenology}",
	journal = {\apjl},
	year = 2011,
	month = feb,
	volume = {727},
	number = {2},
	eid = {L51},
	pages = {L51},
	doi = {10.1088/2041-8205/727/2/L51},
	archivePrefix = {arXiv},
	eprint = {1101.1098},
	primaryClass = {astro-ph.HE},
	adsurl = {https://ui.adsabs.harvard.edu/abs/2011ApJ...727L..51P}
}

@ARTICLE{thom95,
	author = {{Thompson}, Christopher and {Duncan}, Robert C.},
	title = "{The soft gamma repeaters as very strongly magnetized neutron stars - I. Radiative mechanism for outbursts}",
	journal = {\mnras},
	year = 1995,
	month = jul,
	volume = {275},
	number = {2},
	pages = {255-300},
	doi = {10.1093/mnras/275.2.255},
	adsurl = {https://ui.adsabs.harvard.edu/abs/1995MNRAS.275..255T}
}

@ARTICLE{thom01,
	author = {{Thompson}, C. and {Duncan}, R.~C.},
	title = "{The Giant Flare of 1998 August 27 from SGR 1900+14. II. Radiative Mechanism and Physical Constraints on the Source}",
	journal = {\apj},
	year = 2001,
	month = nov,
	volume = {561},
	number = {2},
	pages = {980-1005},
	doi = {10.1086/323256},
	archivePrefix = {arXiv},
	eprint = {astro-ph/0110675},
	primaryClass = {astro-ph},
	adsurl = {https://ui.adsabs.harvard.edu/abs/2001ApJ...561..980T}
}

@ARTICLE{eich89,
	author = {{Eichler}, David and {Livio}, Mario and {Piran}, Tsvi and {Schramm}, David N.},
	title = "{Nucleosynthesis, neutrino bursts and {\ensuremath{\gamma}}-rays from coalescing neutron stars}",
	journal = {\nat},
	year = 1989,
	month = jul,
	volume = {340},
	number = {6229},
	pages = {126-128},
	doi = {10.1038/340126a0},
	adsurl = {https://ui.adsabs.harvard.edu/abs/1989Natur.340..126E}
}

@ARTICLE{woo93,
	author = {{Woosley}, S.~E.},
	title = "{Gamma-Ray Bursts from Stellar Mass Accretion Disks around Black Holes}",
	journal = {\apj},
	year = 1993,
	month = mar,
	volume = {405},
	pages = {273},
	doi = {10.1086/172359},
	adsurl = {https://ui.adsabs.harvard.edu/abs/1993ApJ...405..273W}
}

@ARTICLE{nara92,
	author = {{Narayan}, Ramesh and {Paczynski}, Bohdan and {Piran}, Tsvi},
	title = "{Gamma-Ray Bursts as the Death Throes of Massive Binary Stars}",
	journal = {\apjl},
	year = 1992,
	month = aug,
	volume = {395},
	pages = {L83},
	doi = {10.1086/186493},
	archivePrefix = {arXiv},
	eprint = {astro-ph/9204001},
	primaryClass = {astro-ph},
	adsurl = {https://ui.adsabs.harvard.edu/abs/1992ApJ...395L..83N}
}

@ARTICLE{pacz86,
	author = {{Paczynski}, B.},
	title = "{Gamma-ray bursters at cosmological distances}",
	journal = {\apjl},
	year = 1986,
	month = sep,
	volume = {308},
	pages = {L43-L46},
	doi = {10.1086/184740},
	adsurl = {https://ui.adsabs.harvard.edu/abs/1986ApJ...308L..43P}
}

@ARTICLE{gala98,
	author = {{Galama}, T.~J. and {Vreeswijk}, P.~M. and {van Paradijs}, J. and {Kouveliotou}, C. and {Augusteijn}, T. and {B{\"o}hnhardt}, H. and {Brewer}, J.~P. and {Doublier}, V. and {Gonzalez}, J.-F. and {Leibundgut}, B. and et al.},
	title = "{An unusual supernova in the error box of the {\ensuremath{\gamma}}-ray burst of 25 April 1998}",
	journal = {\nat},
	year = 1998,
	month = oct,
	volume = {395},
	number = {6703},
	pages = {670-672},
	doi = {10.1038/27150},
	archivePrefix = {arXiv},
	eprint = {astro-ph/9806175},
	primaryClass = {astro-ph},
	adsurl = {https://ui.adsabs.harvard.edu/abs/1998Natur.395..670G}
}

@ARTICLE{hior03,
	author = {{Hjorth}, Jens and {Sollerman}, Jesper and {M{\o}ller}, Palle and {Fynbo}, Johan P.~U. and {Woosley}, Stan E. and {Kouveliotou}, Chryssa and {Tanvir}, Nial R. and {Greiner}, Jochen and {Andersen}, Michael I. and {Castro-Tirado}, Alberto J. and et al.},
	title = "{A very energetic supernova associated with the {\ensuremath{\gamma}}-ray burst of 29 March 2003}",
	journal = {\nat},
	year = 2003,
	month = jun,
	volume = {423},
	number = {6942},
	pages = {847-850},
	doi = {10.1038/nature01750},
	archivePrefix = {arXiv},
	eprint = {astro-ph/0306347},
	primaryClass = {astro-ph},
	adsurl = {https://ui.adsabs.harvard.edu/abs/2003Natur.423..847H}
}

@ARTICLE{pacz91,
	author = {{Paczynski}, Bohdan},
	title = "{Cosmological gamma-ray bursts.}",
	journal = {\actaa},
	year = 1991,
	month = jan,
	volume = {41},
	pages = {257-267},
	adsurl = {https://ui.adsabs.harvard.edu/abs/1991AcA....41..257P}
}

@ARTICLE{stan03,
	author = {{Stanek}, K.~Z. and {Matheson}, T. and {Garnavich}, P.~M. and {Martini}, P. and {Berlind}, P. and {Caldwell}, N. and {Challis}, P. and {Brown}, W.~R. and {Schild}, R. and {Krisciunas}, K. and et al.},
	title = "{Spectroscopic Discovery of the Supernova 2003dh Associated with GRB 030329}",
	journal = {\apjl},
	year = 2003,
	month = jul,
	volume = {591},
	number = {1},
	pages = {L17-L20},
	doi = {10.1086/376976},
	archivePrefix = {arXiv},
	eprint = {astro-ph/0304173},
	primaryClass = {astro-ph},
	adsurl = {https://ui.adsabs.harvard.edu/abs/2003ApJ...591L..17S}
}

@ARTICLE{palm05,
	author = {{Palmer}, D.~M. and {Barthelmy}, S. and {Gehrels}, N. and {Kippen}, R.~M. and {Cayton}, T. and {Kouveliotou}, C. and {Eichler}, D. and {Wijers}, R.~A.~M.~J. and {Woods}, P.~M. and {Granot}, J. and et al.},
	title = "{A giant {\ensuremath{\gamma}}-ray flare from the magnetar SGR 1806 - 20}",
	journal = {\nat},
	year = 2005,
	month = apr,
	volume = {434},
	number = {7037},
	pages = {1107-1109},
	doi = {10.1038/nature03525},
	archivePrefix = {arXiv},
	eprint = {astro-ph/0503030},
	primaryClass = {astro-ph},
	adsurl = {https://ui.adsabs.harvard.edu/abs/2005Natur.434.1107P}
}

@ARTICLE{hur05,
	author = {{Hurley}, K. and {Boggs}, S.~E. and {Smith}, D.~M. and {Duncan}, R.~C. and {Lin}, R. and {Zoglauer}, A. and {Krucker}, S. and {Hurford}, G. and {Hudson}, H. and {Wigger}, C. and et al.},
	title = "{An exceptionally bright flare from SGR 1806-20 and the origins of short-duration {\ensuremath{\gamma}}-ray bursts}",
	journal = {\nat},
	year = 2005,
	month = apr,
	volume = {434},
	number = {7037},
	pages = {1098-1103},
	doi = {10.1038/nature03519},
	archivePrefix = {arXiv},
	eprint = {astro-ph/0502329},
	primaryClass = {astro-ph},
	adsurl = {https://ui.adsabs.harvard.edu/abs/2005Natur.434.1098H}
}

@ARTICLE{gott23,
	author = {{Gottlieb}, Ore and {Metzger}, Brian D. and {Quataert}, Eliot and {Issa}, Danat and {Martineau}, Tia and {Foucart}, Francois and {Duez}, Matthew D. and {Kidder}, Lawrence E. and {Pfeiffer}, Harald P. and {Scheel}, Mark A.},
	title = "{A Unified Picture of Short and Long Gamma-Ray Bursts from Compact Binary Mergers}",
	journal = {\apjl},
	year = 2023,
	month = dec,
	volume = {958},
	number = {2},
	eid = {L33},
	pages = {L33},
	doi = {10.3847/2041-8213/ad096e},
	archivePrefix = {arXiv},
	eprint = {2309.00038},
	primaryClass = {astro-ph.HE},
	adsurl = {https://ui.adsabs.harvard.edu/abs/2023ApJ...958L..33G}
}

@ARTICLE{meng24,
	author = {{Meng}, Yan-Zhi and {Wang}, Xiangyu Ivy and {Liu}, Zi-Ke},
	title = "{Significant Cocoon Emission and Photosphere Duration Stretching in GRB 211211A: A Burst from a Neutron Star{\ensuremath{-}}Black Hole Merger}",
	journal = {\apj},
	year = 2024,
	month = mar,
	volume = {963},
	number = {2},
	eid = {112},
	pages = {112},
	doi = {10.3847/1538-4357/ad1bd7},
	archivePrefix = {arXiv},
	eprint = {2304.00893},
	primaryClass = {astro-ph.HE},
	adsurl = {https://ui.adsabs.harvard.edu/abs/2024ApJ...963..112M}
}

@ARTICLE{svin21,
	author = {{Svinkin}, D. and {Frederiks}, D. and {Hurley}, K. and {Aptekar}, R. and {Golenetskii}, S. and {Lysenko}, A. and {Ridnaia}, A.~V. and {Tsvetkova}, A. and {Ulanov}, M. and {Cline}, T.~L. and et al.},
	title = "{A bright {\ensuremath{\gamma}}-ray flare interpreted as a giant magnetar flare in NGC 253}",
	journal = {\nat},
	year = 2021,
	month = jan,
	volume = {589},
	number = {7841},
	pages = {211-213},
	doi = {10.1038/s41586-020-03076-9},
	archivePrefix = {arXiv},
	eprint = {2101.05104},
	primaryClass = {astro-ph.HE},
	adsurl = {https://ui.adsabs.harvard.edu/abs/2021Natur.589..211S}
}

@ARTICLE{robe21,
	author = {{Roberts}, O.~J. and {Veres}, P. and {Baring}, M.~G. and {Briggs}, M.~S. and {Kouveliotou}, C. and {Bissaldi}, E. and {Younes}, G. and {Chastain}, S.~I. and {DeLaunay}, J.~J. and {Huppenkothen}, D. and et al.},
	title = "{Rapid spectral variability of a giant flare from a magnetar in NGC 253}",
	journal = {\nat},
	year = 2021,
	month = jan,
	volume = {589},
	number = {7841},
	pages = {207-210},
	doi = {10.1038/s41586-020-03077-8},
	archivePrefix = {arXiv},
	eprint = {2101.05146},
	primaryClass = {astro-ph.HE},
	adsurl = {https://ui.adsabs.harvard.edu/abs/2021Natur.589..207R}
}

@ARTICLE{yang20,
	author = {{Yang}, Jun and {Chand}, Vikas and {Zhang}, Bin-Bin and {Yang}, Yu-Han and {Zou}, Jin-Hang and {Yang}, Yi-Si and {Zhao}, Xiao-Hong and {Shao}, Lang and {Xiong}, Shao-Lin and {Luo}, Qi and et al.},
	title = "{GRB 200415A: A Short Gamma-Ray Burst from a Magnetar Giant Flare?}",
	journal = {\apj},
	year = 2020,
	month = aug,
	volume = {899},
	number = {2},
	eid = {106},
	pages = {106},
	doi = {10.3847/1538-4357/aba745},
	archivePrefix = {arXiv},
	eprint = {2010.05128},
	primaryClass = {astro-ph.HE},
	adsurl = {https://ui.adsabs.harvard.edu/abs/2020ApJ...899..106Y}
}

@ARTICLE{maz08,
	author = {{Mazets}, E.~P. and {Aptekar}, R.~L. and {Cline}, T.~L. and {Frederiks}, D.~D. and {Goldsten}, J.~O. and {Golenetskii}, S.~V. and {Hurley}, K. and {von Kienlin}, A. and {Pal'shin}, V.~D.},
	title = "{A Giant Flare from a Soft Gamma Repeater in the Andromeda Galaxy (M31)}",
	journal = {\apj},
	year = 2008,
	month = jun,
	volume = {680},
	number = {1},
	pages = {545-549},
	doi = {10.1086/587955},
	archivePrefix = {arXiv},
	eprint = {0712.1502},
	primaryClass = {astro-ph},
	adsurl = {https://ui.adsabs.harvard.edu/abs/2008ApJ...680..545M}
}

@ARTICLE{fred07b,
	author = {{Frederiks}, D.~D. and {Palshin}, V.~D. and {Aptekar}, R.~L. and {Golenetskii}, S.~V. and {Cline}, T.~L. and {Mazets}, E.~P.},
	title = "{On the possibility of identifying the short hard burst GRB 051103 with a giant flare from a soft gamma repeater in the M81 group of galaxies}",
	journal = {Astronomy Letters},
	year = 2007,
	month = jan,
	volume = {33},
	number = {1},
	pages = {19-24},
	doi = {10.1134/S1063773707010021},
	archivePrefix = {arXiv},
	eprint = {astro-ph/0609544},
	primaryClass = {astro-ph},
	adsurl = {https://ui.adsabs.harvard.edu/abs/2007AstL...33...19F}
}

@ARTICLE{zhang20,
	author = {{Zhang}, Hai-Ming and {Liu}, Ruo-Yu and {Zhong}, Shu-Qing and {Wang}, Xiang-Yu},
	title = "{Magnetar Giant Flare Origin for GRB 200415A Inferred from a New Scaling Relation}",
	journal = {\apjl},
	year = 2020,
	month = nov,
	volume = {903},
	number = {2},
	eid = {L32},
	pages = {L32},
	doi = {10.3847/2041-8213/abc2c9},
	archivePrefix = {arXiv},
	eprint = {2008.05097},
	primaryClass = {astro-ph.HE},
	adsurl = {https://ui.adsabs.harvard.edu/abs/2020ApJ...903L..32Z}
}

@ARTICLE{fred07a,
	author = {{Frederiks}, D.~D. and {Golenetskii}, S.~V. and {Palshin}, V.~D. and {Aptekar}, R.~L. and {Ilyinskii}, V.~N. and {Oleinik}, F.~P. and {Mazets}, E.~P. and {Cline}, T.~L.},
	title = "{Giant flare in SGR 1806-20 and its Compton reflection from the Moon}",
	journal = {Astronomy Letters},
	year = 2007,
	month = jan,
	volume = {33},
	number = {1},
	pages = {1-18},
	doi = {10.1134/S106377370701001X},
	archivePrefix = {arXiv},
	eprint = {astro-ph/0612289},
	primaryClass = {astro-ph},
	adsurl = {https://ui.adsabs.harvard.edu/abs/2007AstL...33....1F}
}

@ARTICLE{ofek08,
	author = {{Ofek}, E.~O. and {Muno}, M. and {Quimby}, R. and {Kulkarni}, S.~R. and {Stiele}, H. and {Pietsch}, W. and {Nakar}, E. and {Gal-Yam}, A. and {Rau}, A. and {Cameron}, P.~B. and et al.},
	title = "{GRB 070201: A Possible Soft Gamma-Ray Repeater in M31}",
	journal = {\apj},
	year = 2008,
	month = jul,
	volume = {681},
	number = {2},
	pages = {1464-1469},
	doi = {10.1086/587686},
	archivePrefix = {arXiv},
	eprint = {0712.3585},
	primaryClass = {astro-ph},
	adsurl = {https://ui.adsabs.harvard.edu/abs/2008ApJ...681.1464O}
}

@ARTICLE{ofek06,
	author = {{Ofek}, E.~O. and {Kulkarni}, S.~R. and {Nakar}, E. and {Cenko}, S.~B. and {Cameron}, P.~B. and {Frail}, D.~A. and {Gal-Yam}, A. and {Soderberg}, A.~M. and {Fox}, D.~B.},
	title = "{The Short-Hard GRB 051103: Observations and Implications for Its Nature}",
	journal = {\apj},
	year = 2006,
	month = nov,
	volume = {652},
	number = {1},
	pages = {507-511},
	doi = {10.1086/507837},
	archivePrefix = {arXiv},
	eprint = {astro-ph/0609582},
	primaryClass = {astro-ph},
	adsurl = {https://ui.adsabs.harvard.edu/abs/2006ApJ...652..507O}
}

@ARTICLE{patel25b,
	author = {{Patel}, Anirudh and {Metzger}, Brian D. and {Cehula}, Jakub and {Burns}, Eric and {Goldberg}, Jared A. and {Thompson}, Todd A.},
	title = "{Direct Evidence for r-process Nucleosynthesis in Delayed MeV Emission from the SGR 1806─20 Magnetar Giant Flare}",
	journal = {\apjl},
	year = 2025,
	month = may,
	volume = {984},
	number = {1},
	eid = {L29},
	pages = {L29},
	doi = {10.3847/2041-8213/adc9b0},
	archivePrefix = {arXiv},
	eprint = {2501.09181},
	primaryClass = {astro-ph.HE},
	adsurl = {https://ui.adsabs.harvard.edu/abs/2025ApJ...984L..29P}
}

@ARTICLE{patel25a,
	author = {{Patel}, Anirudh and {Metzger}, Brian D. and {Goldberg}, Jared A. and {Cehula}, Jakub and {Thompson}, Todd A. and {Renzo}, Mathieu},
	title = "{r-process Nucleosynthesis and Radioactively Powered Transients from Magnetar Giant Flares}",
	journal = {\apj},
	year = 2025,
	month = jun,
	volume = {985},
	number = {2},
	eid = {234},
	pages = {234},
	doi = {10.3847/1538-4357/adceb7},
	archivePrefix = {arXiv},
	eprint = {2501.17253},
	primaryClass = {astro-ph.HE},
	adsurl = {https://ui.adsabs.harvard.edu/abs/2025ApJ...985..234P}
}

@ARTICLE{ras22,
	author = {{Rastinejad}, Jillian C. and {Gompertz}, Benjamin P. and {Levan}, Andrew J. and {Fong}, Wen-fai and {Nicholl}, Matt and {Lamb}, Gavin P. and {Malesani}, Daniele B. and {Nugent}, Anya E. and {Oates}, Samantha R. and {Tanvir}, Nial R. and et al.},
	title = "{A kilonova following a long-duration gamma-ray burst at 350 Mpc}",
	journal = {\nat},
	year = 2022,
	month = dec,
	volume = {612},
	number = {7939},
	pages = {223-227},
	doi = {10.1038/s41586-022-05390-w},
	archivePrefix = {arXiv},
	eprint = {2204.10864},
	primaryClass = {astro-ph.HE},
	adsurl = {https://ui.adsabs.harvard.edu/abs/2022Natur.612..223R}
}

@ARTICLE{zhong24,
	author = {{Zhong}, Shu-Qing and {Li}, Long and {Xiao}, Di and {Sun}, Hui and {Zhang}, Bin-Bin and {Dai}, Zi-Gao},
	title = "{The Very Early Soft X-Ray Plateau of GRB 230307A: Signature of an Evolving Radiative Efficiency in Magnetar Wind Dissipation?}",
	journal = {\apjl},
	year = 2024,
	month = mar,
	volume = {963},
	number = {1},
	eid = {L26},
	pages = {L26},
	doi = {10.3847/2041-8213/ad2852},
	archivePrefix = {arXiv},
	eprint = {2402.10619},
	primaryClass = {astro-ph.HE},
	adsurl = {https://ui.adsabs.harvard.edu/abs/2024ApJ...963L..26Z}
}

@ARTICLE{abbott17b,
	author = {{Abbott}, B.~P. and {Abbott}, R. and {Abbott}, T.~D. and {Acernese}, F. and et al.},	
	title = "{Multi-messenger Observations of a Binary Neutron Star Merger}",
	journal = {\apjl},
	year = 2017,
	month = oct,
	volume = {848},
	number = {2},
	eid = {L12},
	pages = {L12},
	doi = {10.3847/2041-8213/aa91c9},
	archivePrefix = {arXiv},
	eprint = {1710.05833},
	primaryClass = {astro-ph.HE},
	adsurl = {https://ui.adsabs.harvard.edu/abs/2017ApJ...848L..12A}
}

@ARTICLE{abbott17c,
	author = {{Abbott}, B.~P. and {Abbott}, R. and {Abbott}, T.~D. and {Acernese}, F. and {Ackley}, K. and et al.},	
	title = "{Gravitational Waves and Gamma-Rays from a Binary Neutron Star Merger: GW170817 and GRB 170817A}",
	journal = {\apjl},
	year = 2017,
	month = oct,
	volume = {848},
	number = {2},
	eid = {L13},
	pages = {L13},
	doi = {10.3847/2041-8213/aa920c},
	archivePrefix = {arXiv},
	eprint = {1710.05834},
	primaryClass = {astro-ph.HE},
	adsurl = {https://ui.adsabs.harvard.edu/abs/2017ApJ...848L..13A}
}

@ARTICLE{abbott17a,
	author = {{Abbott}, B.~P. and {Abbott}, R. and {Abbott}, T.~D. and {Acernese}, F. and et al.},	
	title = "{GW170817: Observation of Gravitational Waves from a Binary Neutron Star Inspiral}",
	journal = {\prl},
	year = 2017,
	month = oct,
	volume = {119},
	number = {16},
	eid = {161101},
	pages = {161101},
	doi = {10.1103/PhysRevLett.119.161101},
	archivePrefix = {arXiv},
	eprint = {1710.05832},
	primaryClass = {gr-qc},
	adsurl = {https://ui.adsabs.harvard.edu/abs/2017PhRvL.119p1101A}
}

@ARTICLE{arnett82,
	author = {{Arnett}, W.~D.},
	title = "{Type I supernovae. I - Analytic solutions for the early part of the light curve}",
	journal = {\apj},
	year = 1982,
	month = feb,
	volume = {253},
	pages = {785-797},
	doi = {10.1086/159681},
	adsurl = {https://ui.adsabs.harvard.edu/abs/1982ApJ...253..785A}
}

@ARTICLE{bob17,
	author = {{Bobrick}, Alexey and {Davies}, Melvyn B. and {Church}, Ross P.},
	title = "{Mass transfer in white dwarf-neutron star binaries}",
	journal = {\mnras},
	year = 2017,
	month = may,
	volume = {467},
	number = {3},
	pages = {3556-3575},
	doi = {10.1093/mnras/stx312},
	archivePrefix = {arXiv},
	eprint = {1702.02377},
	primaryClass = {astro-ph.HE},
	adsurl = {https://ui.adsabs.harvard.edu/abs/2017MNRAS.467.3556B}
}

@ARTICLE{chatz12,
	author = {{Chatzopoulos}, E. and {Wheeler}, J. Craig and {Vinko}, J.},
	title = "{Generalized Semi-analytical Models of Supernova Light Curves}",
	journal = {\apj},
	year = 2012,
	month = feb,
	volume = {746},
	number = {2},
	eid = {121},
	pages = {121},
	doi = {10.1088/0004-637X/746/2/121},
	archivePrefix = {arXiv},
	eprint = {1111.5237},
	primaryClass = {astro-ph.HE},
	adsurl = {https://ui.adsabs.harvard.edu/abs/2012ApJ...746..121C}
}

@ARTICLE{della06,
	author = {{Della Valle}, M. and {Chincarini}, G. and {Panagia}, N. and {Tagliaferri}, G. and {Malesani}, D. and {Testa}, V. and {Fugazza}, D. and {Campana}, S. and {Covino}, S. and {Mangano}, V. and {Antonelli}, L.~A. and {D'Avanzo}, P. and {Hurley}, K. and {Mirabel}, I.~F. and {Pellizza}, L.~J. and {Piranomonte}, S. and {Stella}, L.},
	title = "{An enigmatic long-lasting {\ensuremath{\gamma}}-ray burst not accompanied by a bright supernova}",
	journal = {\nat},
	year = 2006,
	month = dec,
	volume = {444},
	number = {7122},
	pages = {1050-1052},
	doi = {10.1038/nature05374},
	archivePrefix = {arXiv},
	eprint = {astro-ph/0608322},
	primaryClass = {astro-ph},
	adsurl = {https://ui.adsabs.harvard.edu/abs/2006Natur.444.1050D}
}

@ARTICLE{gal06,
	author = {{Gal-Yam}, A. and {Fox}, D.~B. and {Price}, P.~A. and {Ofek}, E.~O. and {Davis}, M.~R. and {Leonard}, D.~C. and {Soderberg}, A.~M. and {Schmidt}, B.~P. and {Lewis}, K.~M. and {Peterson}, B.~A. and {Kulkarni}, S.~R. and {Berger}, E. and {Cenko}, S.~B. and {Sari}, R. and {Sharon}, K. and {Frail}, D. and {Moon}, D. -S. and {Brown}, P.~J. and {Cucchiara}, A. and {Harrison}, F. and {Piran}, T. and {Persson}, S.~E. and {McCarthy}, P.~J. and {Penprase}, B.~E. and {Chevalier}, R.~A. and {MacFadyen}, A.~I.},
	title = "{A novel explosive process is required for the {\ensuremath{\gamma}}-ray burst GRB 060614}",
	journal = {\nat},
	year = 2006,
	month = dec,
	volume = {444},
	number = {7122},
	pages = {1053-1055},
	doi = {10.1038/nature05373},
	archivePrefix = {arXiv},
	eprint = {astro-ph/0608257},
	primaryClass = {astro-ph},
	adsurl = {https://ui.adsabs.harvard.edu/abs/2006Natur.444.1053G}
}

@ARTICLE{kasen10,
	author = {{Kasen}, Daniel and {Bildsten}, Lars},
	title = "{Supernova Light Curves Powered by Young Magnetars}",
	journal = {\apj},
	year = 2010,
	month = jul,
	volume = {717},
	number = {1},
	pages = {245-249},
	doi = {10.1088/0004-637X/717/1/245},
	archivePrefix = {arXiv},
	eprint = {0911.0680},
	primaryClass = {astro-ph.HE},
	adsurl = {https://ui.adsabs.harvard.edu/abs/2010ApJ...717..245K}
}

@ARTICLE{kasen17,
	author = {{Kasen}, Daniel and {Metzger}, Brian and {Barnes}, Jennifer and {Quataert}, Eliot and {Ramirez-Ruiz}, Enrico},
	title = "{Origin of the heavy elements in binary neutron-star mergers from a gravitational-wave event}",
	journal = {\nat},
	year = 2017,
	month = nov,
	volume = {551},
	number = {7678},
	pages = {80-84},
	doi = {10.1038/nature24453},
	archivePrefix = {arXiv},
	eprint = {1710.05463},
	primaryClass = {astro-ph.HE},
	adsurl = {https://ui.adsabs.harvard.edu/abs/2017Natur.551...80K}
}

@ARTICLE{lilong20,
	author = {{Li}, Long and {Dai}, Zi-Gao and {Wang}, Shan-Qin and {Zhong}, Shu-Qing},
	title = "{On the Energy Sources of the Most Luminous Supernova ASASSN-15lh}",
	journal = {\apj},
	year = 2020,
	month = sep,
	volume = {900},
	number = {2},
	eid = {121},
	pages = {121},
	doi = {10.3847/1538-4357/aba95b},
	archivePrefix = {arXiv},
	eprint = {2007.13464},
	primaryClass = {astro-ph.HE},
	adsurl = {https://ui.adsabs.harvard.edu/abs/2020ApJ...900..121L}
}

@ARTICLE{lv22,
	author = {{L{\"u}}, Hou-Jun and {Yuan}, Hao-Yu and {Yi}, Ting-Feng and {Wang}, Xiang-Gao and {Hu}, You-Dong and {Yuan}, Yong and {Rice}, Jared and {Wang}, Jian-Guo and {Cao}, Jia-Xin and {Kong}, De-Feng and {Fernandez-Garc{\'\i}a}, Emilio and {Castro-Tirado}, Alberto J. and {Lian}, Ji-Shun and {Gan}, Wen-Pei and {Wang}, Shan-Qin and {Xin}, Li-Ping and {Caballero-Garc{\'\i}a}, M.~D. and {Fan}, Yu-Feng and {Liang}, En-Wei},
	title = "{GRB 211227A as a Peculiar Long Gamma-Ray Burst from a Compact Star Merger}",
	journal = {\apjl},
	year = 2022,
	month = jun,
	volume = {931},
	number = {2},
	eid = {L23},
	pages = {L23},
	doi = {10.3847/2041-8213/ac6e3a},
	archivePrefix = {arXiv},
	eprint = {2201.06395},
	primaryClass = {astro-ph.HE},
	adsurl = {https://ui.adsabs.harvard.edu/abs/2022ApJ...931L..23L}
}

@ARTICLE{met12,
	author = {{Metzger}, B.~D.},
	title = "{Nuclear-dominated accretion and subluminous supernovae from the merger of a white dwarf with a neutron star or black hole}",
	journal = {\mnras},
	year = 2012,
	month = jan,
	volume = {419},
	number = {1},
	pages = {827-840},
	doi = {10.1111/j.1365-2966.2011.19747.x},
	archivePrefix = {arXiv},
	eprint = {1105.6096},
	primaryClass = {astro-ph.HE},
	adsurl = {https://ui.adsabs.harvard.edu/abs/2012MNRAS.419..827M}
}

@ARTICLE{woo10,
	author = {{Woosley}, S.~E.},
	title = "{Bright Supernovae from Magnetar Birth}",
	journal = {\apjl},
	year = 2010,
	month = aug,
	volume = {719},
	number = {2},
	pages = {L204-L207},
	doi = {10.1088/2041-8205/719/2/L204},
	archivePrefix = {arXiv},
	eprint = {0911.0698},
	primaryClass = {astro-ph.HE},
	adsurl = {https://ui.adsabs.harvard.edu/abs/2010ApJ...719L.204W}
}

@ARTICLE{yang22,
	author = {{Yang}, Jun and {Ai}, Shunke and {Zhang}, Bin-Bin and {Zhang}, Bing and {Liu}, Zi-Ke and {Wang}, Xiangyu Ivy and {Yang}, Yu-Han and {Yin}, Yi-Han and {Li}, Ye and {L{\"u}}, Hou-Jun},
	title = "{A long-duration gamma-ray burst with a peculiar origin}",
	journal = {\nat},
	year = 2022,
	month = dec,
	volume = {612},
	number = {7939},
	pages = {232-235},
	doi = {10.1038/s41586-022-05403-8},
	archivePrefix = {arXiv},
	eprint = {2204.12771},
	primaryClass = {astro-ph.HE},
	adsurl = {https://ui.adsabs.harvard.edu/abs/2022Natur.612..232Y}
}

@ARTICLE{zen19,
	author = {{Zenati}, Yossef and {Perets}, Hagai B. and {Toonen}, Silvia},
	title = "{Neutron star-white dwarf mergers: early evolution, physical properties, and outcomes}",
	journal = {\mnras},
	year = 2019,
	month = jun,
	volume = {486},
	number = {2},
	pages = {1805-1813},
	doi = {10.1093/mnras/stz316},
	archivePrefix = {arXiv},
	eprint = {1807.09777},
	primaryClass = {astro-ph.HE},
	adsurl = {https://ui.adsabs.harvard.edu/abs/2019MNRAS.486.1805Z}
}

@ARTICLE{zen20,
	author = {{Zenati}, Yossef and {Bobrick}, Alexey and {Perets}, Hagai B.},
	title = "{Faint rapid red transients from neutron star-CO white dwarf mergers}",
	journal = {\mnras},
	year = 2020,
	month = apr,
	volume = {493},
	number = {3},
	pages = {3956-3965},
	doi = {10.1093/mnras/staa507},
	archivePrefix = {arXiv},
	eprint = {1908.10866},
	primaryClass = {astro-ph.SR},
	adsurl = {https://ui.adsabs.harvard.edu/abs/2020MNRAS.493.3956Z}
}

@ARTICLE{zhu22,
	author = {{Zhu}, Jin-Ping and {Wang}, Xiangyu Ivy and {Sun}, Hui and {Yang}, Yuan-Pei and {Li}, Zhuo and {Hu}, Rui-Chong and {Qin}, Ying and {Wu}, Shichao},
	title = "{Long-duration Gamma-Ray Burst and Associated Kilonova Emission from Fast-spinning Black Hole-Neutron Star Mergers}",
	journal = {\apjl},
	year = 2022,
	month = sep,
	volume = {936},
	number = {1},
	eid = {L10},
	pages = {L10},
	doi = {10.3847/2041-8213/ac85ad},
	archivePrefix = {arXiv},
	eprint = {2207.10470},
	primaryClass = {astro-ph.HE},
	adsurl = {https://ui.adsabs.harvard.edu/abs/2022ApJ...936L..10Z}
}

@ARTICLE{gao22,
	author = {{Gao}, He and {Lei}, Wei-Hua and {Zhu}, Zi-Pei},
	title = "{GRB 211211A: a Prolonged Central Engine under a Strong Magnetic Field Environment}",
	journal = {\apjl},
	year = 2022,
	month = jul,
	volume = {934},
	number = {1},
	eid = {L12},
	pages = {L12},
	doi = {10.3847/2041-8213/ac80c7},
	archivePrefix = {arXiv},
	eprint = {2205.05031},
	primaryClass = {astro-ph.HE},
	adsurl = {https://ui.adsabs.harvard.edu/abs/2022ApJ...934L..12G}
}

@ARTICLE{met14b,
	author = {{Metzger}, Brian D. and {Piro}, Anthony L.},
	title = "{Optical and X-ray emission from stable millisecond magnetars formed from the merger of binary neutron stars}",
	journal = {\mnras},
	year = 2014,
	month = apr,
	volume = {439},
	number = {4},
	pages = {3916-3930},
	doi = {10.1093/mnras/stu247},
	archivePrefix = {arXiv},
	eprint = {1311.1519},
	primaryClass = {astro-ph.HE},
	adsurl = {https://ui.adsabs.harvard.edu/abs/2014MNRAS.439.3916M}
}

@ARTICLE{piro11,
	author = {{Piro}, Anthony L. and {Ott}, Christian D.},
	title = "{Supernova Fallback onto Magnetars and Propeller-powered Supernovae}",
	journal = {\apj},
	year = 2011,
	month = aug,
	volume = {736},
	number = {2},
	eid = {108},
	pages = {108},
	doi = {10.1088/0004-637X/736/2/108},
	archivePrefix = {arXiv},
	eprint = {1104.0252},
	primaryClass = {astro-ph.HE},
	adsurl = {https://ui.adsabs.harvard.edu/abs/2011ApJ...736..108P}
}

@ARTICLE{xiao24,
	author = {{Xiao}, Shuo and {Zhang}, Yan-Qiu and {Zhu}, Zi-Pei and {Xiong}, Shao-Lin and {Gao}, He and {Xu}, Dong and {Zhang}, Shuang-Nan and {Peng}, Wen-Xi and {Li}, Xiao-Bo and {Zhang}, Peng and et al.},
	title = "{The Peculiar Precursor of a Gamma-Ray Burst from a Binary Merger Involving a Magnetar}",
	journal = {\apj},
	year = 2024,
	month = jul,
	volume = {970},
	number = {1},
	eid = {6},
	pages = {6},
	doi = {10.3847/1538-4357/ad4ee1},
	archivePrefix = {arXiv},
	eprint = {2205.02186},
	primaryClass = {astro-ph.HE},
	adsurl = {https://ui.adsabs.harvard.edu/abs/2024ApJ...970....6X}
}

@ARTICLE{zhang25,
	author = {{Zhang}, Bing},
	title = "{On the duration of gamma-ray bursts}",
	journal = {Journal of High Energy Astrophysics},
	year = 2025,
	month = mar,
	volume = {45},
	pages = {325-332},
	doi = {10.1016/j.jheap.2024.12.013},
	archivePrefix = {arXiv},
	eprint = {2501.00239},
	primaryClass = {astro-ph.HE},
	adsurl = {https://ui.adsabs.harvard.edu/abs/2025JHEAp..45..325Z}
}

@ARTICLE{yi25a,
	author = {{Yi}, S.-X. and {Wang}, C.-W. and {Shao}, X. and {Moradi}, R. and {Gao}, H. and {Zhang}, B. and {Xiong}, S.-L. and {Zhang}, S.-N. and {Tan}, W.-J. and {Liu}, J.-C. and et al.},
	title = "{Evidence of Minijet Emission in a Large Emission Zone from a Magnetically Dominated Gamma-Ray Burst Jet}",
	journal = {\apj},
	year = 2025,
	month = jun,
	volume = {985},
	number = {2},
	eid = {239},
	pages = {239},
	doi = {10.3847/1538-4357/adcf98},
	archivePrefix = {arXiv},
	eprint = {2310.07205},
	primaryClass = {astro-ph.HE},
	adsurl = {https://ui.adsabs.harvard.edu/abs/2025ApJ...985..239Y}
}

@ARTICLE{yi25b,
	author = {{Yi}, Shu-Xu and {Yorgancioglu}, Emre Seyit and {Xiong}, S.-L. and {Zhang}, S.-N.},
	title = "{Long pulse by short central engine: Prompt emission from expanding dissipation rings in the jet front of gamma-ray bursts}",
	journal = {Journal of High Energy Astrophysics},
	year = 2025,
	month = jul,
	volume = {47},
	eid = {100359},
	pages = {100359},
	doi = {10.1016/j.jheap.2025.100359},
	archivePrefix = {arXiv},
	eprint = {2411.16174},
	primaryClass = {astro-ph.HE},
	adsurl = {https://ui.adsabs.harvard.edu/abs/2025JHEAp..4700359Y}
}

@ARTICLE{liu25,
	author = {{Liu}, Xiao-Xuan and {L{\"u}}, Hou-Jun and {Chen}, Qiu-Hong and {Du}, Zhao-Wei and {Liang}, En-Wei},
	title = "{Neutron Star─White Dwarf Merger as One Possible Optional Source of Kilonova-like Emission: Implications for GRB 211211A}",
	journal = {\apjl},
	year = 2025,
	month = aug,
	volume = {988},
	number = {2},
	eid = {L46},
	pages = {L46},
	doi = {10.3847/2041-8213/adec83},
	archivePrefix = {arXiv},
	eprint = {2507.04318},
	primaryClass = {astro-ph.HE},
	adsurl = {https://ui.adsabs.harvard.edu/abs/2025ApJ...988L..46L}
}

@ARTICLE{suv22,
	author = {{Suvorov}, A.~G. and {Kuan}, H.~J. and {Kokkotas}, K.~D.},
	title = "{Quasi-periodic oscillations in precursor flares via seismic aftershocks from resonant shattering}",
	journal = {\aap},
	year = 2022,
	month = aug,
	volume = {664},
	eid = {A177},
	pages = {A177},
	doi = {10.1051/0004-6361/202244082},
	archivePrefix = {arXiv},
	eprint = {2205.11112},
	primaryClass = {astro-ph.HE},
	adsurl = {https://ui.adsabs.harvard.edu/abs/2022A&A...664A.177S}
}

@ARTICLE{sari98,
	author = {{Sari}, Re'em and {Piran}, Tsvi and {Narayan}, Ramesh},
	title = "{Spectra and Light Curves of Gamma-Ray Burst Afterglows}",
	journal = {\apjl},
	year = 1998,
	month = apr,
	volume = {497},
	number = {1},
	pages = {L17-L20},
	doi = {10.1086/311269},
	archivePrefix = {arXiv},
	eprint = {astro-ph/9712005},
	primaryClass = {astro-ph},
	adsurl = {https://ui.adsabs.harvard.edu/abs/1998ApJ...497L..17S}
}

@ARTICLE{huang99,
	author = {{Huang}, Y.~F. and {Dai}, Z.~G. and {Lu}, T.},
	title = "{A generic dynamical model of gamma-ray burst remnants}",
	journal = {\mnras},
	year = 1999,
	month = oct,
	volume = {309},
	number = {2},
	pages = {513-516},
	doi = {10.1046/j.1365-8711.1999.02887.x},
	archivePrefix = {arXiv},
	eprint = {astro-ph/9906370},
	primaryClass = {astro-ph},
	adsurl = {https://ui.adsabs.harvard.edu/abs/1999MNRAS.309..513H}
}

@ARTICLE{evans09,
	author = {{Evans}, P.~A. and {Beardmore}, A.~P. and {Page}, K.~L. and {Osborne}, J.~P. and {O'Brien}, P.~T. and {Willingale}, R. and {Starling}, R.~L.~C. and {Burrows}, D.~N. and {Godet}, O. and {Vetere}, L. and {Racusin}, J. and {Goad}, M.~R. and {Wiersema}, K. and {Angelini}, L. and {Capalbi}, M. and {Chincarini}, G. and {Gehrels}, N. and {Kennea}, J.~A. and {Margutti}, R. and {Morris}, D.~C. and {Mountford}, C.~J. and {Pagani}, C. and {Perri}, M. and {Romano}, P. and {Tanvir}, N.},
	title = "{Methods and results of an automatic analysis of a complete sample of Swift-XRT observations of GRBs}",
	journal = {\mnras},
	year = 2009,
	month = aug,
	volume = {397},
	number = {3},
	pages = {1177-1201},
	doi = {10.1111/j.1365-2966.2009.14913.x},
	archivePrefix = {arXiv},
	eprint = {0812.3662},
	primaryClass = {astro-ph},
	adsurl = {https://ui.adsabs.harvard.edu/abs/2009MNRAS.397.1177E}
}

@ARTICLE{rom05,
	author = {{Roming}, Peter W.~A. and {Kennedy}, Thomas E. and {Mason}, Keith O. and {Nousek}, John A. and {Ahr}, Lindy and {Bingham}, Richard E. and {Broos}, Patrick S. and {Carter}, Mary J. and {Hancock}, Barry K. and {Huckle}, Howard E. and {Hunsberger}, S.~D. and {Kawakami}, Hajime and {Killough}, Ronnie and {Koch}, T. Scott and {McLelland}, Michael K. and {Smith}, Kelly and {Smith}, Philip J. and {Soto}, Juan Carlos and {Boyd}, Patricia T. and {Breeveld}, Alice A. and {Holland}, Stephen T. and {Ivanushkina}, Mariya and {Pryzby}, Michael S. and {Still}, Martin D. and {Stock}, Joseph},
	title = "{The Swift Ultra-Violet/Optical Telescope}",
	journal = {\ssr},
	year = 2005,
	month = oct,
	volume = {120},
	number = {3-4},
	pages = {95-142},
	doi = {10.1007/s11214-005-5095-4},
	archivePrefix = {arXiv},
	eprint = {astro-ph/0507413},
	primaryClass = {astro-ph},
	adsurl = {https://ui.adsabs.harvard.edu/abs/2005SSRv..120...95R}
}

@ARTICLE{cou17,
	author = {{Coulter}, D.~A. and {Foley}, R.~J. and {Kilpatrick}, C.~D. and {Drout}, M.~R. and {Piro}, A.~L. and {Shappee}, B.~J. and {Siebert}, M.~R. and {Simon}, J.~D. and {Ulloa}, N. and {Kasen}, D. and {Madore}, B.~F. and {Murguia-Berthier}, A. and {Pan}, Y. -C. and {Prochaska}, J.~X. and {Ramirez-Ruiz}, E. and {Rest}, A. and {Rojas-Bravo}, C.},
	title = "{Swope Supernova Survey 2017a (SSS17a), the optical counterpart to a gravitational wave source}",
	journal = {Science},
	year = 2017,
	month = dec,
	volume = {358},
	number = {6370},
	pages = {1556-1558},
	doi = {10.1126/science.aap9811},
	archivePrefix = {arXiv},
	eprint = {1710.05452},
	primaryClass = {astro-ph.HE},
	adsurl = {https://ui.adsabs.harvard.edu/abs/2017Sci...358.1556C}
}

@ARTICLE{gom14,
	author = {{Gompertz}, B.~P. and {O'Brien}, P.~T. and {Wynn}, G.~A.},
	title = "{Magnetar powered GRBs: explaining the extended emission and X-ray plateau of short GRB light curves}",
	journal = {\mnras},
	year = 2014,
	month = feb,
	volume = {438},
	number = {1},
	pages = {240-250},
	doi = {10.1093/mnras/stt2165},
	archivePrefix = {arXiv},
	eprint = {1311.1505},
	primaryClass = {astro-ph.HE},
	adsurl = {https://ui.adsabs.harvard.edu/abs/2014MNRAS.438..240G}
}

@ARTICLE{ill75,
	author = {{Illarionov}, A.~F. and {Sunyaev}, R.~A.},
	title = "{Why the Number of Galactic X-ray Stars Is so Small?}",
	journal = {\aap},
	year = 1975,
	month = feb,
	volume = {39},
	pages = {185},
	adsurl = {https://ui.adsabs.harvard.edu/abs/1975A&A....39..185I}
}

@ARTICLE{mei22,
	author = {{Mei}, Alessio and {Banerjee}, Biswajit and {Oganesyan}, Gor and {Salafia}, Om Sharan and {Giarratana}, Stefano and {Branchesi}, Marica and {D'Avanzo}, Paolo and {Campana}, Sergio and {Ghirlanda}, Giancarlo and {Ronchini}, Samuele and {Shukla}, Amit and {Tiwari}, Pawan},
	title = "{Gigaelectronvolt emission from a compact binary merger}",
	journal = {\nat},
	year = 2022,
	month = dec,
	volume = {612},
	number = {7939},
	pages = {236-239},
	doi = {10.1038/s41586-022-05404-7},
	archivePrefix = {arXiv},
	eprint = {2205.08566},
	primaryClass = {astro-ph.HE},
	adsurl = {https://ui.adsabs.harvard.edu/abs/2022Natur.612..236M}
}

@ARTICLE{tro17,
	author = {{Troja}, E. and {Piro}, L. and {van Eerten}, H. and {Wollaeger}, R.~T. and {Im}, M. and {Fox}, O.~D. and {Butler}, N.~R. and {Cenko}, S.~B. and {Sakamoto}, T. and {Fryer}, C.~L. and {Ricci}, R. and {Lien}, A. and {Ryan}, R.~E. and {Korobkin}, O. and {Lee}, S. -K. and {Burgess}, J.~M. and {Lee}, W.~H. and {Watson}, A.~M. and {Choi}, C. and {Covino}, S. and {D'Avanzo}, P. and {Fontes}, C.~J. and {Gonz{\'a}lez}, J. Becerra and {Khandrika}, H.~G. and {Kim}, J. and {Kim}, S. -L. and {Lee}, C. -U. and {Lee}, H.~M. and {Kutyrev}, A. and {Lim}, G. and {S{\'a}nchez-Ram{\'\i}rez}, R. and {Veilleux}, S. and {Wieringa}, M.~H. and {Yoon}, Y.},
	title = "{The X-ray counterpart to the gravitational-wave event GW170817}",
	journal = {\nat},
	year = 2017,
	month = nov,
	volume = {551},
	number = {7678},
	pages = {71-74},
	doi = {10.1038/nature24290},
	archivePrefix = {arXiv},
	eprint = {1710.05433},
	primaryClass = {astro-ph.HE},
	adsurl = {https://ui.adsabs.harvard.edu/abs/2017Natur.551...71T}
}

@ARTICLE{tro22,
	author = {{Troja}, E. and {Fryer}, C.~L. and {O'Connor}, B. and {Ryan}, G. and {Dichiara}, S. and {Kumar}, A. and {Ito}, N. and {Gupta}, R. and {Wollaeger}, R.~T. and {Norris}, J.~P. and {Kawai}, N. and {Butler}, N.~R. and {Aryan}, A. and {Misra}, K. and {Hosokawa}, R. and {Murata}, K.~L. and {Niwano}, M. and {Pandey}, S.~B. and {Kutyrev}, A. and {van Eerten}, H.~J. and {Chase}, E.~A. and {Hu}, Y. -D. and {Caballero-Garcia}, M.~D. and {Castro-Tirado}, A.~J.},
	title = "{A nearby long gamma-ray burst from a merger of compact objects}",
	journal = {\nat},
	year = 2022,
	month = dec,
	volume = {612},
	number = {7939},
	pages = {228-231},
	doi = {10.1038/s41586-022-05327-3},
	archivePrefix = {arXiv},
	eprint = {2209.03363},
	primaryClass = {astro-ph.HE},
	adsurl = {https://ui.adsabs.harvard.edu/abs/2022Natur.612..228T}
}

@ARTICLE{zhang22,
	author = {{Zhang}, Hai-Ming and {Huang}, Yi-Yun and {Zheng}, Jian-He and {Liu}, Ruo-Yu and {Wang}, Xiang-Yu},
	title = "{Fermi-LAT Detection of a GeV Afterglow from a Compact Stellar Merger}",
	journal = {\apjl},
	year = 2022,
	month = jul,
	volume = {933},
	number = {1},
	eid = {L22},
	pages = {L22},
	doi = {10.3847/2041-8213/ac7b23},
	archivePrefix = {arXiv},
	eprint = {2205.09675},
	primaryClass = {astro-ph.HE},
	adsurl = {https://ui.adsabs.harvard.edu/abs/2022ApJ...933L..22Z}
}

@ARTICLE{gib17,
	author = {{Gibson}, S.~L. and {Wynn}, G.~A. and {Gompertz}, B.~P. and {O'Brien}, P.~T.},
	title = "{Fallback accretion on to a newborn magnetar: short GRBs with extended emission}",
	journal = {\mnras},
	year = 2017,
	month = oct,
	volume = {470},
	number = {4},
	pages = {4925-4940},
	doi = {10.1093/mnras/stx1531},
	archivePrefix = {arXiv},
	eprint = {1706.04802},
	primaryClass = {astro-ph.HE},
	adsurl = {https://ui.adsabs.harvard.edu/abs/2017MNRAS.470.4925G}
}

@ARTICLE{mor24,
	author = {{Mor{\'a}n-Fraile}, Javier and {R{\"o}pke}, Friedrich K. and {Pakmor}, R{\"u}diger and {Aloy}, Miguel A. and {Ohlmann}, Sebastian T. and {Schneider}, Fabian R.~N. and {Leidi}, Giovanni and {Lioutas}, Georgios},
	title = "{Self-consistent magnetohydrodynamic simulation of jet launching in a neutron star - white dwarf merger}",
	journal = {\aap},
	year = 2024,
	month = jan,
	volume = {681},
	eid = {A41},
	pages = {A41},
	doi = {10.1051/0004-6361/202347555},
	archivePrefix = {arXiv},
	eprint = {2310.08623},
	primaryClass = {astro-ph.HE},
	adsurl = {https://ui.adsabs.harvard.edu/abs/2024A&A...681A..41M}
}

@ARTICLE{wang24,
	author = {{Wang}, Xiangyu Ivy and {Yu}, Yun-Wei and {Ren}, Jia and {Yang}, Jun and {Zou}, Ze-Cheng and {Zhu}, Jin-Ping},
	title = "{What Powered the Kilonova-like Emission after GRB 230307A in the Framework of a Neutron Star─White Dwarf Merger?}",
	journal = {\apjl},
	year = 2024,
	month = mar,
	volume = {964},
	number = {1},
	eid = {L9},
	pages = {L9},
	doi = {10.3847/2041-8213/ad2df6},
	archivePrefix = {arXiv},
	eprint = {2402.11304},
	primaryClass = {astro-ph.HE},
	adsurl = {https://ui.adsabs.harvard.edu/abs/2024ApJ...964L...9W}
}

@ARTICLE{dai98a,
	author = {{Dai}, Z.~G. and {Lu}, T.},
	title = "{Gamma-ray burst afterglows and evolution of postburst fireballs with energy injection from strongly magnetic millisecond pulsars}",
	journal = {\aap},
	year = 1998,
	month = may,
	volume = {333},
	pages = {L87-L90},
	doi = {10.48550/arXiv.astro-ph/9810402},
	archivePrefix = {arXiv},
	eprint = {astro-ph/9810402},
	primaryClass = {astro-ph},
	adsurl = {https://ui.adsabs.harvard.edu/abs/1998A&A...333L..87D}
}

@ARTICLE{dai98b,
	author = {{Dai}, Z.~G. and {Lu}, T.},
	title = "{{\ensuremath{\gamma}}-Ray Bursts and Afterglows from Rotating Strange Stars and Neutron Stars}",
	journal = {\prl},
	year = 1998,
	month = nov,
	volume = {81},
	number = {20},
	pages = {4301-4304},
	doi = {10.1103/PhysRevLett.81.4301},
	archivePrefix = {arXiv},
	eprint = {astro-ph/9810332},
	primaryClass = {astro-ph},
	adsurl = {https://ui.adsabs.harvard.edu/abs/1998PhRvL..81.4301D}
}

@ARTICLE{dich23,
	author = {{Dichiara}, S. and {Tsang}, D. and {Troja}, E. and {Neill}, D. and {Norris}, J.~P. and {Yang}, Y. -H.},
	title = "{A Luminous Precursor in the Extremely Bright GRB 230307A}",
	journal = {\apjl},
	year = 2023,
	month = sep,
	volume = {954},
	number = {1},
	eid = {L29},
	pages = {L29},
	doi = {10.3847/2041-8213/acf21d},
	archivePrefix = {arXiv},
	eprint = {2307.02996},
	primaryClass = {astro-ph.HE},
	adsurl = {https://ui.adsabs.harvard.edu/abs/2023ApJ...954L..29D}
}

@ARTICLE{fyn06,
	author = {{Fynbo}, Johan P.~U. and {Watson}, Darach and {Th{\"o}ne}, Christina C. and {Sollerman}, Jesper and {Bloom}, Joshua S. and {Davis}, Tamara M. and {Hjorth}, Jens and {Jakobsson}, P{\'a}ll and {J{\o}rgensen}, Uffe G. and {Graham}, John F. and {Fruchter}, Andrew S. and {Bersier}, David and {Kewley}, Lisa and {Cassan}, Arnaud and {Castro Cer{\'o}n}, Jos{\'e} Mar{\'\i}a and {Foley}, Suzanne and {Gorosabel}, Javier and {Hinse}, Tobias C. and {Horne}, Keith D. and {Jensen}, Brian L. and {Klose}, Sylvio and {Kocevski}, Daniel and {Marquette}, Jean-Baptiste and {Perley}, Daniel and {Ramirez-Ruiz}, Enrico and {Stritzinger}, Maximilian D. and {Vreeswijk}, Paul M. and {Wijers}, Ralph A.~M. and {Woller}, Kristian G. and {Xu}, Dong and {Zub}, Marta},
	title = "{No supernovae associated with two long-duration {\ensuremath{\gamma}}-ray bursts}",
	journal = {\nat},
	year = 2006,
	month = dec,
	volume = {444},
	number = {7122},
	pages = {1047-1049},
	doi = {10.1038/nature05375},
	archivePrefix = {arXiv},
	eprint = {astro-ph/0608313},
	primaryClass = {astro-ph},
	adsurl = {https://ui.adsabs.harvard.edu/abs/2006Natur.444.1047F}
}

@ARTICLE{geh06,
	author = {{Gehrels}, N. and {Norris}, J.~P. and {Barthelmy}, S.~D. and {Granot}, J. and {Kaneko}, Y. and {Kouveliotou}, C. and {Markwardt}, C.~B. and {M{\'e}sz{\'a}ros}, P. and {Nakar}, E. and {Nousek}, J.~A. and {O'Brien}, P.~T. and {Page}, M. and {Palmer}, D.~M. and {Parsons}, A.~M. and {Roming}, P.~W.~A. and {Sakamoto}, T. and {Sarazin}, C.~L. and {Schady}, P. and {Stamatikos}, M. and {Woosley}, S.~E.},
	title = "{A new {\ensuremath{\gamma}}-ray burst classification scheme from GRB060614}",
	journal = {\nat},
	year = 2006,
	month = dec,
	volume = {444},
	number = {7122},
	pages = {1044-1046},
	doi = {10.1038/nature05376},
	archivePrefix = {arXiv},
	eprint = {astro-ph/0610635},
	primaryClass = {astro-ph},
	adsurl = {https://ui.adsabs.harvard.edu/abs/2006Natur.444.1044G}
}

@ARTICLE{gill23,
	author = {{Gillanders}, James H. and {Troja}, Eleonora and {Fryer}, Chris L. and {Ristic}, Marko and {O'Connor}, Brendan and {Fontes}, Christopher J. and {Yang}, Yu-Han and {Domoto}, Nanae and {Rahmouni}, Salma and {Tanaka}, Masaomi and {Fox}, Ori D. and {Dichiara}, Simone},
	title = "{Heavy element nucleosynthesis associated with a gamma-ray burst}",
	journal = {arXiv e-prints},
	year = 2023,
	month = aug,
	eid = {arXiv:2308.00633},
	pages = {arXiv:2308.00633},
	doi = {10.48550/arXiv.2308.00633},
	archivePrefix = {arXiv},
	eprint = {2308.00633},
	primaryClass = {astro-ph.HE},
	adsurl = {https://ui.adsabs.harvard.edu/abs/2023arXiv230800633G}
}

@ARTICLE{kou93,
	author = {{Kouveliotou}, Chryssa and {Meegan}, Charles A. and {Fishman}, Gerald J. and {Bhat}, Narayana P. and {Briggs}, Michael S. and {Koshut}, Thomas M. and {Paciesas}, William S. and {Pendleton}, Geoffrey N.},
	title = "{Identification of Two Classes of Gamma-Ray Bursts}",
	journal = {\apjl},
	year = 1993,
	month = aug,
	volume = {413},
	pages = {L101},
	doi = {10.1086/186969},
	adsurl = {https://ui.adsabs.harvard.edu/abs/1993ApJ...413L.101K}
}

@ARTICLE{cheong25,
	author = {{Cheong}, Patrick Chi-Kit and {Pitik}, Tetyana and {Longo Micchi}, Lu{\'\i}s Felipe and {Radice}, David},
	title = "{Gamma-Ray Bursts and Kilonovae from the Accretion-induced Collapse of White Dwarfs}",
	journal = {\apjl},
	year = 2025,
	month = jan,
	volume = {978},
	number = {2},
	eid = {L38},
	pages = {L38},
	doi = {10.3847/2041-8213/ada1cc},
	archivePrefix = {arXiv},
	eprint = {2410.10938},
	primaryClass = {astro-ph.HE},
	adsurl = {https://ui.adsabs.harvard.edu/abs/2025ApJ...978L..38C}
}

@ARTICLE{levan24,
	author = {{Levan}, Andrew J. and {Gompertz}, Benjamin P. and {Salafia}, Om Sharan and {Bulla}, Mattia and {Burns}, Eric and {Hotokezaka}, Kenta and {Izzo}, Luca and {Lamb}, Gavin P. and {Malesani}, Daniele B. and {Oates}, Samantha R. and et al.},
	title = "{Heavy-element production in a compact object merger observed by JWST}",
	journal = {\nat},
	year = 2024,
	month = feb,
	volume = {626},
	number = {8000},
	pages = {737-741},
	doi = {10.1038/s41586-023-06759-1},
	archivePrefix = {arXiv},
	eprint = {2307.02098},
	primaryClass = {astro-ph.HE},
	adsurl = {https://ui.adsabs.harvard.edu/abs/2024Natur.626..737L}
}

@ARTICLE{sun25,
	author = {{Sun}, H. and {Wang}, C.-W. and {Yang}, J. and {Zhang}, B.-B. and {Xiong}, S.-L. and {Yin}, Y.-H.~I. and {Liu}, Y. and {Li}, Y. and {Xue}, W.-C. and {Yan}, Z. and et al.},
	title = "{Magnetar emergence in a peculiar gamma-ray burst from a compact star merger}",
	journal = {National Science Review},
	year = 2025,
	month = mar,
	volume = {12},
	number = {3},
	eid = {nwae401},
	pages = {nwae401},
	doi = {10.1093/nsr/nwae401},
	archivePrefix = {arXiv},
	eprint = {2307.05689},
	primaryClass = {astro-ph.HE},
	adsurl = {https://ui.adsabs.harvard.edu/abs/2025NSRev..12E.401S}
}

\end{document}